\documentclass{aa}  

\usepackage{graphicx}
\usepackage{txfonts}
\usepackage{xcolor}
\usepackage{ulem}
\usepackage{multicol}
\usepackage{amsmath}
\usepackage[toc]{appendix}
\usepackage[T1]{fontenc}
\usepackage{ae,aecompl}

\usepackage{newtxtext,newtxmath}
\usepackage{graphicx}	
\usepackage{amsmath}	
\usepackage{amssymb}	

\usepackage{natbib}
\bibpunct{(}{)}{;}{a}{}{,} 

\usepackage{hyperref}

\newcommand{\Msun}{$\rm{M_{\odot}}$}

\newcommand{\afe}{[$\alpha$/Fe]  }

\begin{document}

   \title{Recent star formation episodes in the Galaxy: impact on its chemical properties and the evolution of its abundance gradient}
   \titlerunning{Recent star formation episodes in the Galaxy }
   \author{Tianxiang~Chen
          \inst{1},
          Nikos~Prantzos
          \inst{1}
          }
    \institute{
            Institut d’Astrophysique de Paris, UMR7095 CNRS, Sorbonne Universit\'{e}, 98bis Bd. Arago, 75104 Paris, France\\
            \email{tianxiang.chen@iap.fr}\\
            \email{nikos.prantzos@iap.fr}
             }

\date{Accepted XXX. Received YYY; in original form ZZZ}

\abstract
   {} 
   { We investigate the chemical evolution of the Milky Way disk exploring various schemes of recent (last several Gyr) star formation episodes, as reported in recent observational works.}
   {We use a semi-analytical model with parametrized radial migration and we  introduce gaussian star formation episodes constrained by those recent observations.} 
   { We find significant impact of the star formation episodes on several observables, like the local age-metallicity and \afe \ vs metallicity relations, as well as the local stellar metallicity distribution or the existence of young \afe-rich stars. Moreover, we show that the recently found "wiggly" behaviour of the disk abundance gradient with age can be interpreted in terms of either star formation or infall episodes. }
   {}


   \maketitle


\section{Introduction}
\label{sec:Intro}

For a long time, the Milky Way (MW) was thought as a quiescently evolving disk galaxy for most of its recent history, not really typical of its kind when compared to other galaxies of the same mass \citep[e.g.][]{Hammer_2007}. Reconstructing its past history by counting local star numbers as function of stellar age
was - and remains - difficult because of uncertainties in evaluating stellar ages \citep{Soderblom_2010,Mints_2017,Queiroz_2023}, selection functions and analysis methods \citep[][]{Miglio_2021,Montalban_2021, Anders_2023}. The various releases of Gaia data \citet[][and references therein]{Gaia_DR3}  established that the early (few Gyr) of MW evolution was characterized by intense and episodic star formation (SF) activity, through interactions and mergers with nearby galaxies  \citep[][and reference therein]{Helmi_2020}.

The more recent star formation period of the Galaxy (the last several Gyr) was also scrutinized.
By fitting  main-sequence star \textit{Gaia} data with the Besançon Galaxy Model, \citet{Mor_2019} suggested that a rather extended star formation episode occurred in the MW 2-3 Gyr ago;  \citet{Isern_2019} reached a similar conclusion analyzing the luminosity function of  massive white dwarfs. On the other hand, \citet{RuizLara_2020} obtained the  SF history of stars currently present within a solar-centered  bubble of 2-kpc radius and  found several  enhancements of SF at 5.7, 1.9 and 1 Gyr ago; they identified those SF episodes with the last perigalactic passages of  Sagittarius  dwarf satellite galaxy. Finally,  \citet{Sahlholdt_2022} 
combined data from
 Galactic Archaeology with HERMES \cite[GALAH]{Buder_2021} and \textit{Gaia}, and found two peaks in the recent SF history of the MW, at 2-3 Gyr and 5.7 Gyr ago, not very different from those reported in \citet{RuizLara_2020}.

The evolution of element abundances and abundance ratios  may be strongly affected by the existence of SF episodes or bursts (hereafter SFBs). \citet{Gilmore_1991} studied the implications of the "bursty" SFH of the Small and Large Magellanic Clouds (as identified through colour-magnitude diagrams) and argued that that "if star formation occurs in a short burst and is
followed by a period of negligible star formation, then the
interstellar medium (ISM) will continue to be enriched in
iron - but not in oxygen - by the ejecta of Type I supernovae.
Thus the first stars to form after an extended quiescent period
can have a large underabundance of oxygen relative to iron". Using simple toy models to illustrate their argument, they showed that  
the observed underabundance of [O/Fe] in the
interstellar medium (ISM) of the Magellanic Clouds can be reproduced in a fairy simple way. More recently \citet{Johnson_2020} explored the impact of simple SFBs, driven either by episodic gas accretion or by  episodically enhanced star formation efficiency,  using a one-zone chemical evolution model. Subsequently, \citet{Johnson_2021} introduced a late-SFB - inspired by \citet{Mor_2019} and \citet{Isern_2019} - in their multi-zone galactic chemical evolution model and they found that it could bring a better agreement with the local age-[Fe/H] and age-[O/H] relations observed  by \citet{Feuillet_2019}.

In this paper, we explore various consequences of  the recetnly reported SFBs by \citet{Mor_2019} , \citet{RuizLara_2020} and \citet{Sahlholdt_2022}. We study in some detail the detailed chemical evolution associated with different SFHs based on our updated model of disk galaxy evolution \citep[from][]{Prantzos_2023} which includes stellar radial migration. The observational constraints  are presented in Sect. \ref{sec:OBS}. In Sect. \ref{sec:Model}, we summarize our model and the various kinds of SFBs we introduce. In Sect. \ref{sec:Result} and \ref{sec:Prop_SFB} we present our results and discuss their impact on various observables. In Sec. \label{sec:met_grad_sfb} and \ref{sec:SFB_infall} we discuss in some depth the role of SFBs on the reconstruction of the  evolution of the metallicity gradient at the birth place of stars that are observed locally today, after a method suggested by \citet{Minchev_2018} and elaborated in \citet{LuBuck_2022} and \citet{Ratcliffe_2023}. We summarize our results in Sec. \ref{sec:Discuss}.

\section{Observations}
\label{sec:OBS}
\begin{figure}
	\includegraphics[width=0.42\textwidth]{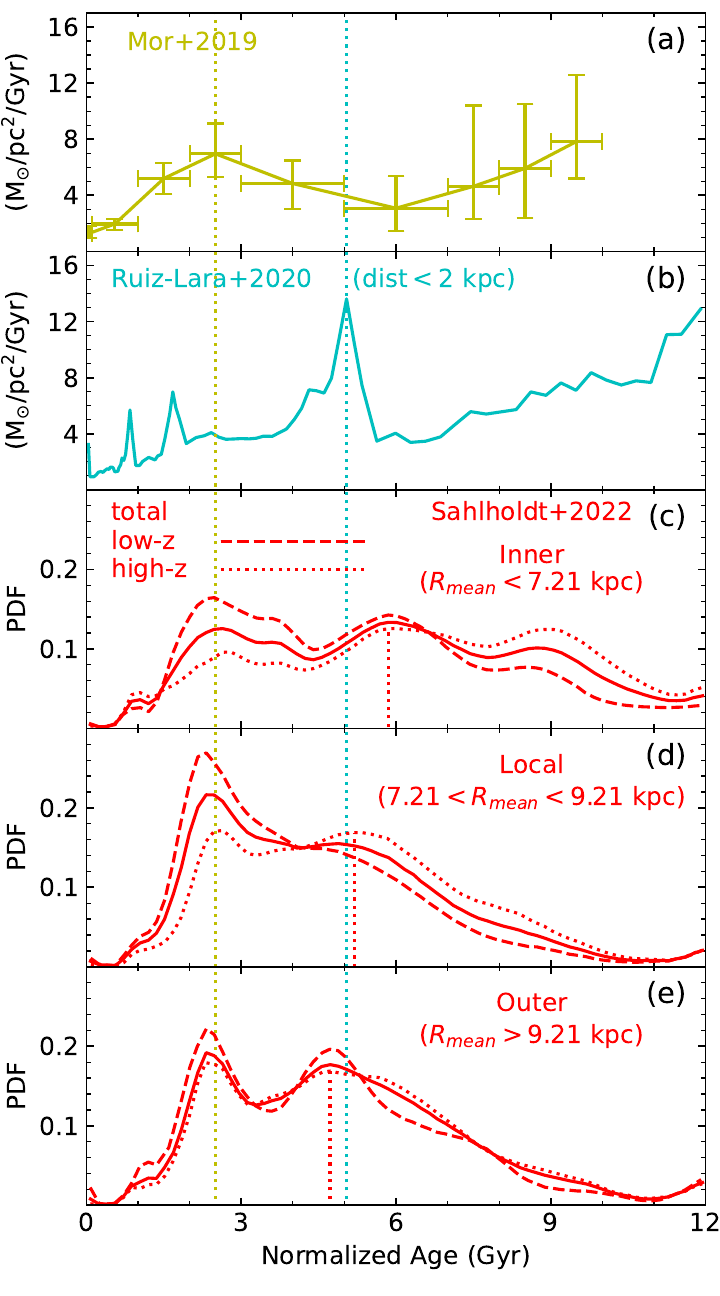}
    \caption{Recently reported episodes of star formation in the late history of the local region of our Galaxy.  {\it From top to bottom}: a) SFR obtained from main-sequence stars \citep{Mor_2019}, with error bars in age and SFR intensity. b) The cyan curve represents the SFH in a bubble of radius $\sim 2$ kpc around the Sun \citep{RuizLara_2020}.  The three subsequent panels display results for age distribution of stars obtained from \citet{Sahlholdt_2022} for (c) the inner, (d) the local and (e) the outer Galactic disk; in those three panels dashed curves correspond to the situation close to the plane and  dotted ones away from the plane, with solid curve representing the total (in precnetage). The age range of all data is normalized to the time duration of our model ($T_{\rm max} =12$ Gyr).  In (a) and (b), observations are normalized to the present day local star surface density (38 \Msun/pc$^2$) with the adopted IMF of \citet{Kroupa_2002}. The two vertical lines are to guide the eye and correspond to the centroids of the SFB of \citep{Mor_2019}, of the strongest SFB of \citep{RuizLara_2020} and of the first SFB of \citet{Sahlholdt_2022}.}
    \label{fig:SFH_OBS}
\end{figure}

Three different sets of observational data \citep{Mor_2019,RuizLara_2020,Sahlholdt_2022} are used to guide our modelling of SFBs in this work. Here we briefly describe their main features and show possible relation in Fig.\ref{fig:SFH_OBS}. 

\subsection{\citet{Mor_2019}}
\label{subsec:Mor}

 The Besançon Galaxy model combined with an approximate Bayesian computation algorithm was used by \citet{Mor_2019} to analyse data from \textit{Gaia} DR2 catalogue, which they estimated to be complete up to magnitude G=12. They explored a parameter space with 15 dimensions that simultaneously includes the initial mass function (IMF) and a non-parametric SFH for the Galactic disk. The SFH displays a decreasing trend at the early age followed by a prominent SFR enhancement beginning around 5 Gyr ago. This enhancement lasted for about 4 Gyr and reached its maximum $\sim$ 2-3 Gyr ago. During that period, about half of the stellar mass currently observed in the solar neighborhhod was formed (Fig. \ref{fig:SFH_OBS} top).

 \cite{Isern_2019} used Gaia data to construct the luminosity function of massive  white dwarfs (0.9 $<$ M/\Msun $<$1.1)in the solar neighborhood, for distances d$<$ 100 pc. Because the lifetime of their progenitors is very short, the birth times of both parent stars and daughter residues are very close and facilitate the reconstruction of an (effective) SFH. \cite{Isern_2019} found that the SFR started growing from zero in the early Galaxy, reached a maximum 6–7 Gyr ago, then declined and $\sim$5 Gy ago started to increase once more, to reach another local maximum 2–3 Gyr ago. The local SFH inferred from the analysis of \cite{Isern_2019} resembles a lot to the one suggested by \cite{Mor_2019}.

\subsection{\citet{RuizLara_2020}}
\label{subsec:Ruiz}
\citet{RuizLara_2020} provided a detailed SFH of the local MW disk by comparing color-magnitude diagrams (CMD) from Gaia DR2 stars within a 2-kpc sphere around the Sun to synthetic ones, computed with BaSTI stellar evolution library \citep{Hidalgo_2018} and IMF from \citet{Kroupa_2002}. The derived SFH shows three short and prominent star forming episodes occurring $\sim$ 5.7, 1.9 and 1 Gyr ago and lasting for about 0.8, 0.2 and 0.1 Gyr, respectively  (Fig. \ref{fig:SFH_OBS} second from top). According to \citet{RuizLara_2020} these episodes seem to coincide with the reported pericenter passages of the Sagittarius (Sgr) dwarf satellite galaxy, as recovered from its reconstructed orbital history. The color-magnitude diagram (CMD) of the Sgr core also revealed three stellar populations corresponding to these epochs \citep{Siegel_2007}. These findings suggest that both the Milky Way disk and Sgr have been possibly affected by their close interactions in the last few Gyrs.

\subsection{\citet{Sahlholdt_2022}}
\label{subsec:Sahl}
Combining GALAH DR3 and Gaia EDR3 data on  turnoff stars \citet{Sahlholdt_2022} established a detailed map of age-metallicity distribution in the MW, which was then sliced into subsamples by tangential velocity and Galactic position. These subsamples reveal three phases in the SFH of the MW separated by two transitions at the ages of 10 Gyr and 4 to 6 Gyr. In the first phase, kinematically hot and high-alpha stars form in the inner disk. In the second phase, the stellar populations gradually become kinematically colder and have lower \afe \ ratio. At this period, stars form with super-solar metallicities in the inner disk and  with sub-solar metallicities in the outer regions. Finally, the third phase is related to a recent SFB (see (Fig. \ref{fig:SFH_OBS} bottom three panels, for the inner, local and outer disk, respectively).

Despite the different methods and age estimates used in the three studies, one may notice that the youngest SFB of \cite{Sahlholdt_2022} has the same centroid in time than the one of \cite{Mor_2019}, while the older of the three SFBs of \cite{RuizLara_2020} roughly coincides with the SFBs of \cite{Sahlholdt_2022} in the local and outer disks. On the other hand, the duration of the SFBs vary widely between the studies, from a fraction of a Gyr in the case of \cite{RuizLara_2020} to a couple of Gyr in the case of \cite{Mor_2019}. Finally, in the case of \cite{Sahlholdt_2022} it seems that different radial regions experience the oldest  SFB at different times. 
since it occurs earlier in the inner disk than in the local and outer ones, as sketched by the dotted red vertical lines in the bottom three panels of Fig. 1: there is a time delay of $\sim$1.5 Gyr between the peaks of SF in the inner and outer regions, separated by $\sim$2.5 kpc (if peak SF intensities and ages are taken at face value). 

In the next section we present our model and in Sections 4 and 5 we explore the consequences of introducing SFBs constrained by the observations presented in the previous subsections.

\section{Model}
\label{sec:Model}

Our model is a one-dimension "multi-ring" model for the MW, with the rings coupled by stellar radial migration. The model is adapted from \citet{Kubryk_2015a, Kubryk_2015b}, as updated recently in \citet{Prantzos_2023}. Here we summarize the main features of the model and present in detail our simple method of introducing episodes of enhanced star formation.

\subsection{Gas Infall and star formation}
\label{subsec:Infall}
We assume that the MW disk is gradually built by primordial gas infall in the potential well of a dark matter halo of $10^{12}$ \Msun. We define the disk as the region outside galactocentric radius R$_{\rm G}$=2 kpc. The formation of the MW disk proceeds "inside-out", with the time scale of gas infall $\tau_{in}$=0.5 Gyr in the innermost zones and gradually increasing with radius up to 7.5 Gyr at 21 kpc.

The "Kennicutt-Schmidt" law \citep{Schmidt_1959,Schmidt_1963,Kennicutt_1998} is widely used to describe the correlation between surface densities of star formation $\Psi$ and gas $\Sigma_{G}$: 
\begin{equation}
\Psi \propto \Sigma_{G}^{N}
\label{eq:KSlaw}
\end{equation}
 
However, \citet{Bigiel_2008}, analyzing data from nearby disk galaxies, found  that the SFR is directly connected to the surface density of molecular gas H$_2$.  Also, based on  updated observations, \citet{Krumholz_2014} suggested that $\rm{H_2}$ correlates better with star formation than the atomic gas HI or the total gas surface density. Following these studies, we assume that the SFR is proportial to $\rm{H_2}$ surface density. The semi-empirical method of \citet{BR_2006} is adopted to calculate the fraction of molecular gas \citep[see Appendix B in][]{Kubryk_2015a} and we calculate the SFR in the  standard model - with no SFB -  as: 
\begin{equation}
\Psi(R,t)=\alpha \Sigma_{H_2}(R,t)
\label{eq:SFR_STD}
\end{equation}

We adopt the Initial Mass Function (IMF) of \citet{Kroupa_2002} with the slope of high mass range equal to 1.3. 

\subsection{Stellar migration}
\label{subsec:StelMigr}
The disk is divided in concentric rings of radial width $\Delta R $=0.5 kpc, with stars allowed to migrate during their evolution.
Following \citet{Kubryk_2015a}, we adopt a statistical prescription to parameterize the epicyclic motion of stars (blurring) and true variation in their guiding radius(churning) separately. We use a time-dependent radial velocity dispersion $\sigma_r \propto \tau^{\beta}$ to describe blurring with  $\beta$=0.25 \citep{Aumer_2016} and we use the method  described in \citet{Kubryk_2015a} for churning. We note that a recent analysis of the kinematics and orbits of 23795 turnoff and giant stars from Gaia-DR2 \citep{Beraldo_2021}, finds that in the Solar neighbourhood, about half of the old thin disc stars can be classified as migrators, while for the thick disc this migrating fraction could be as high as $\sim$1/3. We also note that 
\cite{Feltzing2020} investigated the amount of radial migration in the MW disc, using data for red giant branch stars from APOGEE DR14, parallaxes from Gaia, and stellar ages based on the C and N abundances and results of the models of  \cite{Minchev_2018}, \cite{Frankel2018}, \cite{Sanders2015} and \cite{Kubryk_2015a}. They found that  half of the stars have experienced some sort of radial migration, 10 per cent likely
have suffered only from churning, and a modest 5–7 per cent have never experienced either churning or blurring. They also found that their results depend little on the radial abundance profiles of the adopted models, despite the differences among the latter.

Regarding the locally observed two-branch behaviour of [alpha/Fe] vs metallicity, two classes of evolution, either secular or episodic one have been proposed. Observations suggest that stars with distinctly different histories
(one corresponding to intense early star formation and another occuring more slowly at late times) co-exist locally today. In the “episodic” case both histories occur locally, but they are separated by more or less long
episodes of intense infall and/or paucity in star formation; radial migration plays little or no role in that scheme. In the “secular evolution case” the intense star formation occurs in the inner disk early on, at an epoch where few
stars are formed in the solar vicinity (because of the inside-out formation of the disk) and some of those early stars are transported locally by radial migration, which plays a key role in that case; see discussion in the
Introduction of \cite{Prantzos_2023}.

\subsection{Nucleosynthesis prescriptions}
\label{subsec:Yields}
 
As  in  Prantzos+2023, we use the yields of \citet{Cristallo_2015} for Low and Intermediate Mass Stars (LIMS) and those  of \citet{LimongiChieffi_2018} for Massive Stars\footnote{
\href{http://fruity.oa-abruzzo.inaf.it} {\textcolor{blue}{{http://fruity.oa-abruzzo.inaf.it}}} and \href{http://orfeo.iaps.inaf.it/index.html}{\textcolor{blue}{{http://orfeo.iaps.inaf.it/index.html}}} for LIMS and massive stars, respectively.}, and we include all isotopes from H to U and all relevant nucleosynthesis process and sites.  The yields of \citet{LimongiChieffi_2018} consider mass loss and rotation and we adopt here an Initial Distribution of Rotational Velocities (IDROV) from \cite {Prantzos_2018, Prantzos_2020}, tailored to match several obervables.  \citet{Philcox_2018} found that this prescription provides a better fit than other sets of yields to  the solar composition.

\subsection{Star Formation Episodes}
\label{subsec:SFB}

Episodes of enhanced  star formation may result from a temporary increase of the amount of gas accreted on a galaxy, or of the star formation efficiency after some external perturbation, or from a  combination of the  two. Here we assume  enhanced star formation efficiency, defined as SFR per unit mass of molecular gas rather than total gas, in view of the adopted SFR in Eq. \ref{eq:SFR_STD}. For the standard model, we assume that the star formation efficiency keeps the same value all the time ($\varepsilon$=$\alpha$). In the case of a star formation history (SFH) with episodes of enhanced star formation, we keep the same prescription for infall and we we adopt a modified star formation law 
\begin{equation}
\Psi(R,t)=\alpha \Sigma_{H_2}(R,t)(1+\sum{I_b(R,t)})
\label{eq:SFR_SFB}
\end{equation}
where $I_b(R,t)$ represent episodes of enhanced star formation. We call them "star formation bursts" (SFB), although their intensity and duration may vary considerably. 
In the numerical implementation we assume that a SFB is characterized by a temporal  enhancement of the star formation rate, without any other modification in the code. This would correspond e.g. to a perturbation of the gas by an external agent, like a nearby small galaxy. If the enhancement is assumed to be due to a temporal increase of the infall rate, or to a gas-rich merger, then the effect of the SFB on the chemical composition would be less important, since the enhanced amount of ejected  metals would be diluted to a larger amount of (accreted) gas. On the other hand, one may envisage more complex situations, i.e. the SFB heating the gas and "quenching" star formation for sometime. The real situation would be, undoubtedly, more complicated than the simplified ones explored here.

In this work, we consider three types of SFB:  local ones (occurring in some delimited region of Galactocentric radius), global ones (large perturbations occurring simultaneously over the largest part of the disk) and  propagated SFBs, starting at some place and propagating outwards.

\subsubsection{Local SFB}
In that case, SFB occurs in a small range of Galactocentric radii and we use a multivariate Gaussian function of time and radius to describe this type of burst
\begin{equation}
I_b(R,t)=\gamma_b\exp[{-\frac{(R-\mu_R)^2}{2\sigma_{R}^2}-\frac{(t-\mu_t)^2}{2\sigma_{t}^2}}]
\label{eq:localSFB}
\end{equation}
where $\gamma_b$ is used to describe the maximum intensity of the SFB,  $\mu_t$ is the time of maximum star formation efficiency, $\mu_R$ is the radius where maximum star formation efficiency occurs.  $\sigma_{t}$ is related to the duration of SFB, and $\sigma_{R}$ describes  the radial range affected by the enhanced star formation. Here we adopt for all local bursts ($\mu_{R}$=8 kpc) a fixed radial extension $\sigma_R$=1 kpc, and  we simply adjust  the temporal extension $\mu_{t}$ and the maximal intensity $\gamma_b$ to reproduce as closely as possible the observations.

\subsubsection{Global SFB}
To describe a global SFB affecting the whole disk,  we adopt for all radial zones the same amplitude of enhanced star formation $\gamma_b$ at the same time. This is of course an extreme simplification of the situation, since there is no reason why the SF efficiency should be enhanced by the same factor everywhere. However, the star formation still depends on the total amount of local molecular gas (see Eq. \ref{eq:SFR_SFB}). On the other hand, there are no observational data allowing us to distinguish between local and global SFBs, so we study here the simplest case. The SFB is then simply described by a Gaussian function in time: 
\begin{equation}
I_b(R,t)=\gamma_b\exp[{-\frac{(t-\mu_t)^2}{2\sigma_{t}^2}}]
\label{eq:globalSFB}
\end{equation}
We used both assumptions of local SFB and global SFB to simulate the SFHs constrained by \citet{Mor_2019} (single local burst or single global burst) and \citet{RuizLara_2020} (3 local SFBs or 3 global SFBs) as we discuss in the next section.

\subsubsection{Propagated SFB}

Motivated by the results of \citet{Sahlholdt_2022}, we define a more specific case of  propagated SFB. In that case, we assume that SFB affects delimited radial zones like the local SFB, but with a time delay depending on radius, thus simulating the episode of star formation propagated across the disk. The propagated SFB is then defined by a modified multivariate Gaussian function:
\begin{equation}
I_b(R,t)=\gamma_b\exp[{-\frac{(R-\mu_R)^2}{2\sigma_{R}^2}-\frac{(t-\mu_t- t_R R)^2}{2\sigma_{t}^2}}]
\label{eq:insideoutSFB}
\end{equation}
where $t_R$ is the characteristic timescale (in Gyr/kpc) of radial displacement of the SF episode, i.e. the inverse of the speed of radial displacement. A positive  $t_R$ implies that the SFB occurs earlier in the inner regions and its peak moves outward.

\section{Results}
\label{sec:Result}

\subsection{Star formation}
\label{subsec:SFR}

\begin{figure*}
    \begin{center}      
	\includegraphics[width=0.9\textwidth]{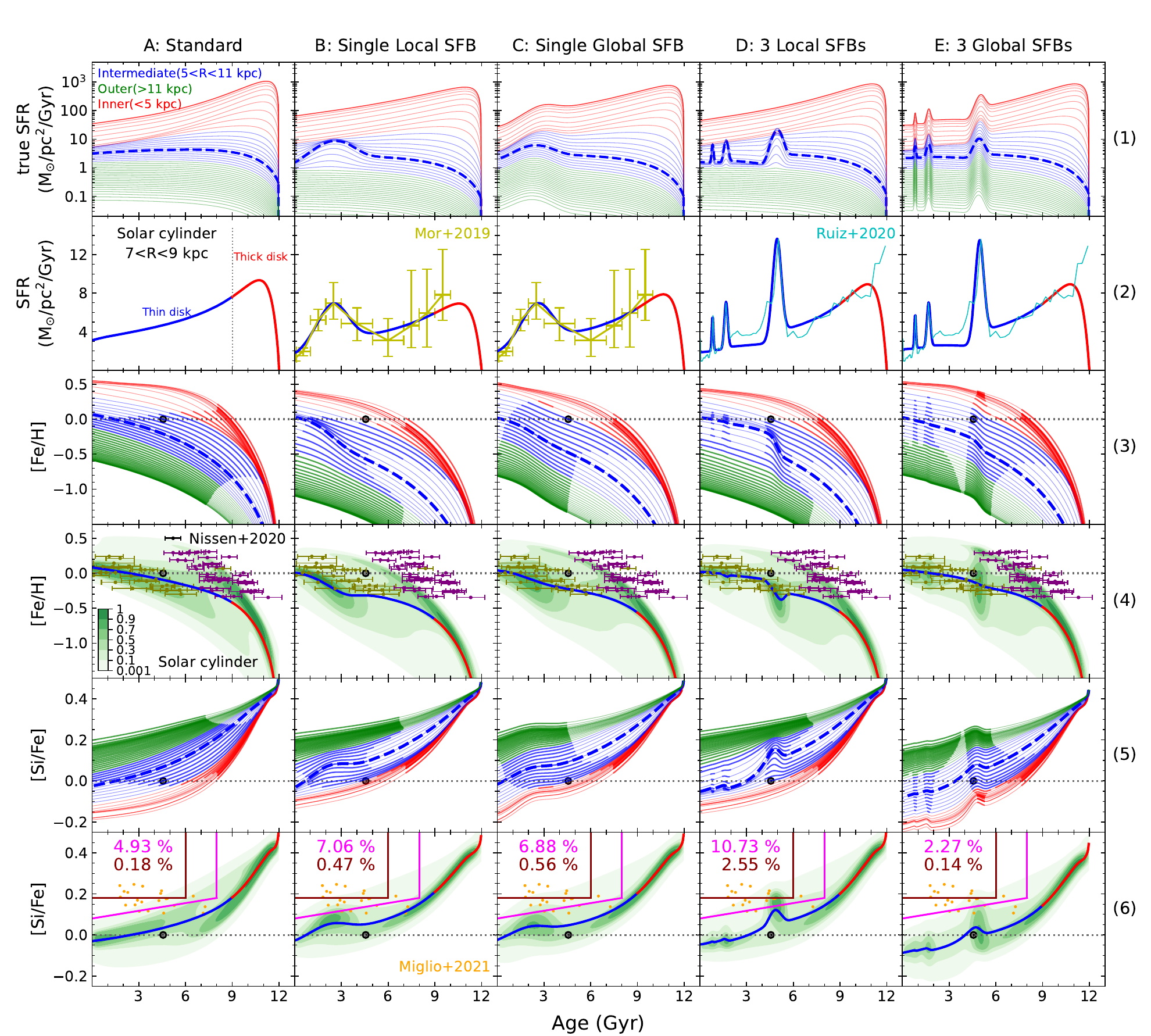}
    \caption{Comparison of various quantities as function of age among models with five different SFHs (standard, single local SFB, single global SFB, 3 local SFBs, 3 global SFBs). 
    $\it Row \: 1$: The true SFR in different radial zones, divided into inner disk ($R<5 $ kpc, red), intermediate disk ($5<R<11$ kpc, blue) and outer disk ($R>11$ kpc, green). The evolution at $R_{\odot}=8$ kpc is depicted by dashed blue curve. 
    $\it Row \: 2$: SFH as observed today in solar cylinder ($7<R<9$ kpc) after counting stellar radial migration. The vertical dotted line at 9 Gyr separates thin disk ($\textup{age}<9$ Gyr, red) from thick disk ($\textup{age}>9$ Gyr, blue). 
    $\it Row \: 3$: Evolution of [Fe/H] in the gas in different radial zones. Thick portions of curves indicate the periods of SFR higher than the time-average one.  
    $\it Row \: 4$: [Fe/H] of stars existing in solar cylinder at the end of the simulation. The color-coded filled contours represent number density of stars. As shown on the left colorbar, the contour levels correspond to the fractions of 0.001, 0.1, 0.3, 0.5, 0.7, 0.9 of total numbers of stars. The red and blue curve indicate average [Fe/H] of thin and thick disk. Orange dots denote 
    young, high \afe stars from \citet{Miglio_2021}. 
    $\it Row \: 5$: Evolution of [Si/Fe] in the gas in different radial zones. 
    $\it Row \: 6$: [Si/Fe] in stars found in the solar cylinder at the end of the simulation. The fraction of young $\alpha$-enhanced stars selected using the criterion from \citet{Chiappini_2014} (magenta) and \citet{ZhangM_2021} (brown) are shown on the top left, along with the corresponding fractions of stars in those regions as obtained by our models (see text).}
    \label{fig:Age__SFR_FeH_XFe}
     \end{center}
\end{figure*}

Fig. \ref{fig:Age__SFR_FeH_XFe} summarizes our results for the cases of  local and global SFBs. In the column A we display some results of the baseline model, which is essentially the same as the one presented in \citet{Prantzos_2023}. In the top left panel (A1), the true SFHs of different radial zones are displayed color-coded, for the  inner disk (Galactocentric radius $R<5$ kpc, red), intermediate ($5<R<11$ kpc, blue) and outer ($R<11$ kpc, green), while the local one   at $R=8$ is depicted by dashed blue curve. The inside-out formation of the disk is clearly seen by the early peak of SFR in the inner zones, compared to the practically constant  local  SFR and the lately increasing SFR of the outer zones.  

In the second  row we display the age distribution of local stars, as seen today by an observer in the solar cylinder ($7<R<9$ kpc). The youngest stars are mostly formed locally, because there is little time for radial migration to redistribute stars over the disk. As one moves to higher ages though, more and more locally formed stars have left and more and more stars produced elsewhere have moved in the local volume. This is especially true for stars produced in the inner disk, where the true SFR has been quite intense early on (see panel A1). As a result, the perceived local SFH is more peaked early on than it was in reality (panel A2). As in \cite{Prantzos_2023}, we assume here  that the old disk (age $>9$ Gyr) corresponds to the thick disk and the  younger one to the thin disk (age $<9$ Gyr), but these considerations are not of importance in this study.

The next two columns in Fig. \ref{fig:Age__SFR_FeH_XFe} display results for a single SFB, either a local (Col. B) or a global one (Col. C). The intensity and duration of the SFB is adjusted as to reproduce in both cases the observationnally inferred SF history in the solar neighborhood by \citet{Mor_2019}, as shown in the 2nd row (panels B2 and C2, respectively).  
Finally, the two columns on the right display results for three SFBs, either local (Col. D) or global ones (Col. E). Their intensity is adjusted as to reproduce  the SFBs inferred for the local disks from the analysis of \cite{RuizLara_2020}, as seen in panels D2 and
E2, respectively). In all cases, fitting the observationally inferred SFBs is the main constrain of our SFB parametrization.

Compared to our baseline model, the introduction of late episodes of enhanced SF,  obviously reduces the importance of the oldest stellar population observed locally, because a larger fraction of stars is made at later times.
It can also be noticed that,  in order to reproduce the same local SF history,  a local SFB has to be more intense than a global one. The reason is that the latter involves more radial zones and thus produces more stars of a given age that migrate and are found in the solar neighborhood today.

\subsection{Abundances vs age }
\label{subsec:AMR}

The two subsequent rows of Fig. \ref{fig:Age__SFR_FeH_XFe} describe the evolution of metallicity in the gas of all the zones (Row 3) and in the stars now present in the solar vicinity (Row 4). Thick parts of the curves in Row 3 correspond more active  periods,  when the SFR in each zone is higher than the corresponding time-average value. These results have been analysed in detail in \citet{Prantzos_2023}.
We  note that in the baseline model, the Sun is found to be formed a couple of kpc inwards from its present position,  at a galactocentric distance R$\sim$6 kpc.  That brings in agreement the fact that solar metallicity is very close to the one of local young stars \citep{Nieva2012} and the ISM \citep{Cartledge2006,Ritchey2023}, while  models suggest a sizeable metallicity evolution in the local gas in the past few Gyr \citep[e.g.][]{Minchev_2013,Kubryk_2015a,Prantzos_2018,Prantzos_2023}

Moving to the right in the Row 3 of Fig. \ref{fig:Age__SFR_FeH_XFe}, we find that the single SFB adjusted as to reproduce the SF history of \citet{Mor_2019}, drives a strong increase of the late metallicity  after Sun's formation (panels B3 and C3), since  the peak of the SFE is located at $\sim$2.5 Gyr in the past. 
Two factors contribute to that increase: a) the enhanced rate of SNIa - the main Fe producer -  a few hundred Myr after the SFB and  b) the small amount of gas left after the intense SFB. The combined result of (a) and (b) is that a large amount of Fe is released in a small amount of gas, rapidly increasing its metallicity. We also note that this late metallicity increase is more pronounced in the case of a local burst (panel B4 ) than in a global one (panel C4). The reason, as explained also in the last paragraph of the previous section, is that the local burst has to be more intense than the global one to reproduce the same current stellar density in the solar vicinity;  as a result, it increases more the local gas metallicity. On the other hand, the first of the 3 SFBs inferred by \cite{RuizLara_2020} to occur 6 Gyr ago lasts much less than the one of \cite{Mor_2019} and produces a steeper increase in [Fe/H] (panels D3 and E3 in Fig. \ref{fig:Age__FeH_Rad})

Our results for the impact of a SFB on the age-metallicity relation differ with the ones presented by \cite{Johnson_2021}. They find that   [Fe/H] is considerably reduced during the SFB, because they assume that the SGB is fuelled by an intense infall which dilutes metallicity. We assume instead that the SFBs are due to perturbations induced by closeby interactions with other galaxies and thus the metallicity in all radial zones of the disk never decreases in the models presented here. We also tested cases with enhanced infall and found  similar dilution effects, but only in the case of local SFBs (see Sec. \ref{sec:SFB_infall}).

\subsubsection{Impact on Sun's birthplace}
\label{subsubsec:SolBirthPlace}

The strong late metallicity evolution induced by the recent SFB has an important impact on the determination of the birth radius of the Sun. Since the present day metallicity of the local gas must be near solar\footnote{In all the models, the star formation efficiency is adjusted so as to lead to a $\sim$solar metallicity in the local ISM today.}, its difference with the local gas metallicity 4.5 Gyr ago is much smaller in the baseline model ($\sim$0.25 dex) than in the SFB senarii B or C (local or global SFB, respectively). For that reason, the Sun's birthplace in that case is in a region further inwards than in the baseline model, around R$\sim$ 4.5 - 5 kpc. The exact values depend, of course on the adopted model and radial migration scheme. However, the key point is that a late SFB (after Sun's formation) "pushes" further inwards the birth position of the Sun, more than a smooth recent SF history.

The situation is slightly different in the case of an SFB prior to the formation of the Sun, as illustrated in panel D3 and E3 of Fig. \ref{fig:Age__SFR_FeH_XFe}, obtained for the SF history of \citet{RuizLara_2020}. The oldest and strongest of the 3 SFBs occurs just prior to the formation of the Sun, increasing abruptly the gas metallicity. After that, star formation activity becomes low (because little gas is left) and metallicity increases very little,  since the two most recent SFBs of \citet{RuizLara_2020} are not strong enough to change that situation. As a result the final local gas metallicity is only $\sim 0.15$ dex  
higher than its was 4.5 Gy ago at R $\sim 8$ kpc. In contrast to the previous case (of a SFB posterior to Sun's birth), the abundance gradient in the intermediate disk is now small, as seen by the density of the blue curves at an age $\sim$ 4.5 Gyr. Despite this small metallicity difference, we note that the birth place of the Sun suggested by the model is still at R$\sim$5 - 5.5 kpc, i.e. slightly smaller than in the baseline model.

Thus, we find that a strong SFB may play an important role in the determination of Sun's birth place. It may increase or decrease the absolute value of the abundance gradient (depending on whether it occurs after or before the Sun's formation, respectively), but it always places the Sun's birth radius further in the inner disk than in the baseline model with no SFB.

We note that the recent study of \cite{LuMinchev_2022} finds a low value for the birth radius of the Sun of R$_{\odot,b}$= 4.5$\pm$0.4 kpc, substantially lower than most other studies, including this one. In their Fig. 6 (top right panel) they display the age-metallicity relations at birth radii, showing a rapid metallicity increase in most radii at approximately the time of Sun's formation, similar to the features displayed in the 3d row of our Fig.  \ref{fig:Age__SFR_FeH_XFe}.

\subsubsection{Overdensities in the plane of abundances vs age}
\label{subsubsec:Overdensities_vs_age}

The fourth row of Fig. \ref{fig:Age__SFR_FeH_XFe} displays the age-[Fe/H] distribution of stars found today in the solar cylinder. The scatter of [Fe/H] at a given age results from stars born in different radial zones with different chemical evolution, as already found in the pioneer paper of \cite{Sellwood_2002}. In our standard model (panel A4) there are two overdense regions, corresponding to a young (a few Gy)  and an old (more than 9 Gyr) stellar population, which are identified to the thin and thick disks. Most stars in the old overdensity are formed early in the inner region and their [Fe/H] increases very rapidly with time. The young overdensity represent stars formed mostly locally, with [Fe/H]  $\sim 0$ and increasing little with time. 

In that same panel A4, we overplot the data of \cite{Nissen2020} who used HARPS spectra to determine metallicities (in the range of -0.3$<$ [Fe/H] $<$+0.3) for 72 nearby solar-type stars and ages bu adopting ASTEC stellar models and
Gaia DR2 parallaxes. They found that their  sample is rather clearly split in two sequences: a sequence of old stars with a steep rise of [Fe/H] to +0.3 dex at an age of $\sim$7 Gyr and a younger sequence with [Fe/H] increasing from about -0.3 dex to +0.2 dex over the last 6 Gyr. They interpreted this split as evidence of two episodes of accretion of gas onto the Galactic disk with a quenching of star formation in between. We find that in our standard model this trend is naturally explained, as can be seen in panel A4, and also explained in detail in \cite[see Fig. 4 in that work]{Prantzos_2023}. The reason is that, in the framework of that model,  in the inner disk (which is older because of the inside-out formation scheme) stars are produced abundantly  and metallicity rises rapidly in the early stages; in contrast, in the local disk metallicity rises slowly and later. Radial migration
brings in the solar vicinity stars from the inner disk and thus two branches (overdensities) are obtained. \cite{Nissen2020} interpret their finding in terms of the two-infall model, in which all the action takes place in the same environment: a 1-zone model with a unique, albeit discontinuous history having two distinct epochs of star formation separated by a hiatus or quenching. In contrast, in multi-zone models with radial migration, the early, intense phase of stellar activity occurs in the inner disk and the late phase mostly locally.

In the 4th row, the single late burst models (local or global) preserve the positions of these two populations, while enhancing the density of the young population with respect to the standard model (panels B4 and C4, to be compared to A4). On the other hand, the 3 local SFBs model (panels D4 and E4) create 3 small and narrow overdensities. 
The first of the 3 local SFBs  (5 Gyr ago) creates a small decrease in the average [Fe/H] vs age relation of the stars found today in the solar cylinder, for ages between 4 and 5 Gyr. This "depression" is due to a complex interplay between the star formation activity of the inner Fe-rich zones prior to the SFB (with some of their stars now migrated at R=8 kpc) and the fact that just after the burst the local ISM is enriched in Fe only by the CCSN, while a few hundreds of Myr later the SNIa of the SFB produce a rapid increase of the Fe abundance. Unsurprisingly, this feature is absent in the case of the global burst (panel E4).

This peculiar feature of a SFB (the delayed overproduction of Fe from SNIa with respect to the $\alpha$-element produced exclusively by CCSN) is also reflected in another observable, namely, the behaviour of \afe \ vs [Fe/H], as seen in the next two rows of Fig. \ref{fig:Age__SFR_FeH_XFe}. 
In the fifth row we show the evolution of [Si/Fe], using Si as a representative  $\alpha$ element. The ratio of CCSN to SNIa 
decreases with time everywhere, but more strongly in the inner disk where the SFR decreases very rapidly after the first couple of Gyr and much less in the outer disk where the SFR remains important till today (see first row in Fig. 
\ref{fig:Age__SFR_FeH_XFe}). As a result, in the standard model (panel A5),  [Si/Fe] of the gaseous disk  decreases steadily, and more rapidly in the inner disk than in the outer one in the early times. In that model, the [Si/Fe] of local stars today (panel A6) decreases rapidly in the earliest times (since most of the local oldest stars have migrated from the inner disk) and then very little (since the youngest  stars are formed mostly locally). Those features are also encountered in the other columns of the last row, but the introduction of SFBs produces different overdensities in the phase space at the epochs of the SFEs. Those overdensities appear at the age of starburst but produce in general higher \afe values than those of  the standard model. This is clearly seen in particular for the local bursts of panels B6 and D6, respectively, where the post-SFB \afe rises by 0.1 dex and 0.3 dex, respectively. We explore the implication of that effect concerning the issue of young and high \afe stars in the next subsection.

\subsubsection{Young \afe-rich stars?}
\label{subsubsec:Young_afe_rich}

In the case of a broad SFB 3 Gyr ago \citep[as the one claimed in][]{Mor_2019}, either local or global (panels B6 and C6 of Fig. \ref{fig:Age__SFR_FeH_XFe}) the decline of Si/Fe is stopped for some time after the SFB, because of the enhanced Si/Fe contributed by CCSN. But later, because of the Fe injected by the SNIa of the SFB, the Si/Fe decreases rapidly, as seen for the stars formed in the past 2 Gyr. Thus,  a single late SFB is expected to leave a clear, albeit weak, signature of a late decrease in the  \afe vs [FeH] relation.

A stronger signature is obtained in the case of a strong and narrow SFB, like the first of the 3 SFBs of \citet{RuizLara_2020}, as displayed in the panels D6 and E6 in Fig. \ref{fig:Age__SFR_FeH_XFe}. This time, an important temporary increase, by $\sim$0.3 dex, is obtained in the [Si/Fe] ratio right after the first burst 5 Gy ago, and for a short period, because that SFB has a shorter duration than the one of \cite{Mor_2019}.

Thus, a generic feature of {\it narrow} SFBs, independently of their time of occurrence, is a temporal enhancement of the  \afe ratio, obtained for a short time and immediately after the SFB. This confirms the original, intuitive work of \citet{Gilmore_1991}, the recent analytical findings of \citet{Weinberg_2017} and the numerical ones by \citet{Johnson_2020}, albeit only qualitatively. The amplitude of the SFBs considered in those works is  smaller than here, and consequently the enhancement of [O/Fe] is also smaller.

\citet{Johnson_2021} and \cite{Borisov_2022} note that this feature  may be of interest regarding the population of the "young [$\alpha$/Fe]-enhanced" stars reported a few years ago  with data from APOGEE  \citet{Chiappini_2015} and \citet{Martig_2015}. Those stars are younger than 5-6 Gyr and designate thin disk stars, while their \afe \ ratio is around 0.2 dex higher than the typical thin disk stars, closer to those of the thick disk. 
\citet{Jofre_2016} suggested that the properties of those  stars may be explained by mass transfer in close binaries, but \citet{Anders_2017} noticed that this hypothesis  cannot explain
the different number counts in the
inner and outer disc. 

With LAMOST data \citet{ZhangM_2021}  and CorroAPOGEE data \citet{Miglio_2021} argued that the highest than average masses of those stars favor the interaction/merger hypothesis, which would make old and \afe-rich stars - normally encountered in the thick disk - appear younger. These conclusions are supported by recent chemical and kinematic studies of  \citet{Queiroz_2023}, \citet{Cerqui2023} and \citet{Grisoni_2024}  who find that these stars display similar kinematic properties with those of the high-$\alpha$ disk. We notice, however, that \citet{Borisov_2022} reach different conclusions using GALAH data, since they find that their sample of "young [$\alpha$/Fe]-enhanced"  dwarf stars  exhibits lithium abundances similar to those of young [$\alpha$/Fe]-normal dwarfs at the same age and [Fe/H]. 

 We investigate here quantitavely whether a  recent SFB induced by enhanced star formation efficiency, without any additional infall of high-$\alpha$ gas, could create a population of young [$\alpha$/Fe] enhanced stars. The boxes delimited by the two orthogonal (brown) or quasi-orthogonal (magenta)  lines in the panels of the 6th row of Fig. \ref{fig:Age__SFR_FeH_XFe} correspond to the regions which should not contain stars,i.e. relatively young and with high \afe) according to \citet{ZhangM_2021} and \citet{Chiappini_2014}, respectively  (the young alpha-rich stars from the K2 survey in \citet{Grisoni_2024} are also contained within the "forbidden" region of the latter reference). Moreover, we plot in those figures the 25 "overmassive" stars (orange points) with high \afe reported in \citet{Miglio_2021}. The percentages reported inside those boxes display the fractions of our model stars found in the corresponding regions. It can be seen that in our standard model, a few percent of our stars are found in the "forbidden" region of \citet{Chiappini_2014}, but less than 1\% in the more restricted zone of \citet{ZhangM_2021}. 
 
 The introduction of SFBs, in particular the oldest of the three of \citet{RuizLara_2020} (panel D6) increases considerably those fractions, but the youngest stars with the highest \afe \ remain inaccessible to our modelling. The reason is that at such late times, the average  \afe in the local and nearby inner disk (the stars of which have time to radially migrate to the solar neighborhood) is quite low and the small recent SFBs of \citet{RuizLara_2020}  or the strong and extended one of \citet{Mor_2019} cannot increase it to the level of \afe$>0.20$ required by the data. A couple of recent (age <3 Gyr) strong  and short SFBs - as the oldest one in \citet{RuizLara_2020} - could help in that respect, but such events have not been reported. Another possibility could be a "top-heavy" IMF during the SFB, increasing the \afe \ to much larger extent (say 0.3 or 0.45 dex) than the conventional IMF adopted here. We do not attempt here to model such events. We simply note that in those cases, the kinematic properties of those young and \afe-rich  stars - resembling those of the thick disk, as noticed by \citet{Queiroz_2023} and \citet{Cerqui2023} - could  be readily explained by the highly turbulent environment inherent to a SFB.

\subsection{Properties vs metallicity}
\begin{figure*}
    \begin{center}        
	\includegraphics[width=0.9\textwidth]{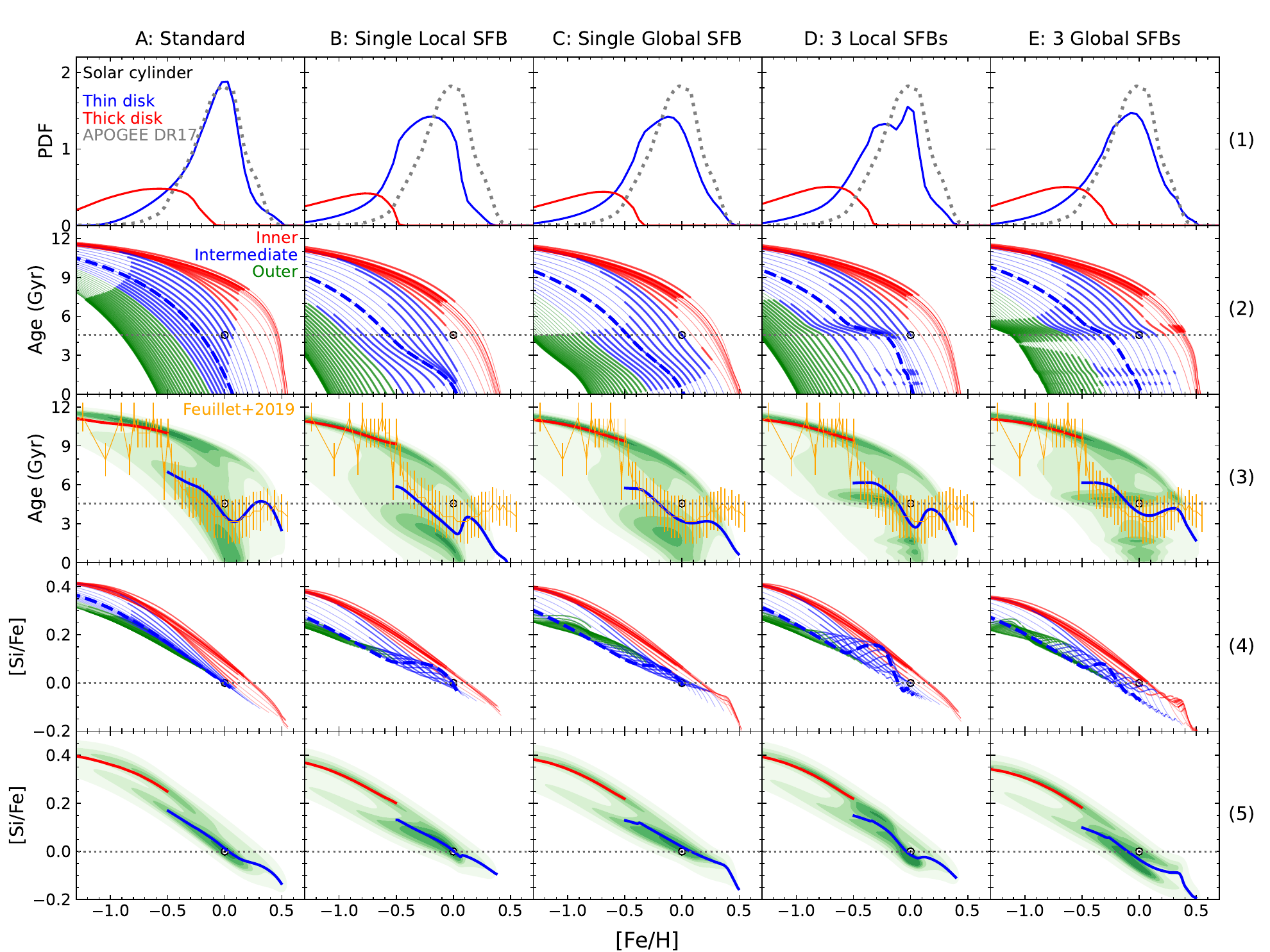}
    \caption{Model results (colour coded as in Fig. 1) and comparison to observations for various quantities as function of [Fe/H]. 
    $\it Row \: 1$: Metallicity distributions for thin (red) and thick (blue) disks in solar cylinder. The thin ones are compared to the red giants near the plane ($|z|<0.3\ \textup{kpc}$) from the APOGEE DR17 (dotted grey).
    $\it Row \: 2$: Age-Metallicity relation color for all radial zones.
    $\it Row \: 3$: Predicted Age vs [Fe/H] distribution present in solar cylinder. Average ages of thin and thick disks (red and blue curves) are compared to data from \citet{Feuillet_2019} shown by orange curves with vertical error bars. 
    $\it Row \: 4$: Gas evolution of the various radial zones of the model in the[Si/Fe]-[Fe/H] plane.
    $\it Row \: 5$: Predicted [Si/Fe] vs [Fe/H] distribution present in solar cylinder.}
    \label{fig:FeH__Hist_Age_XFe}
     \end{center}
\end{figure*}

In Fig. \ref{fig:FeH__Hist_Age_XFe} we present the results of our models as function of metallicity and compare to observational data.

\subsubsection{Metallicity distributions}
\label{subsubsec:MetDistr}
In the top row we display the metallicity distribution (MD) and compare to data for the thin disk (dotted) taken from the 17th data release of APOGEE \citep{APOGEE_DR17}. Following \citet{Hayden_2015},we apply the following restriction and quality criteria to APOGEE DR17:
$1.0\le\log{g}\le3.8$;
$3500\le T_{\textup{eff}}\le5500\ \textup{K}$;
SNR $< 80$;
ASPCAP flag $\notin$ $\texttt{STAR\_BAD}$ or $\texttt{VSINI\_WARN}$

Using distance from Gaia EDR 3 \citep{BJ_2021}, we limit the radial Galactocentric distance to the range $7<R<9 \textup{ kpc}$ and distance from the plane to $|z|<0.3\ \textup{kpc}$ to get a sample of mostly thin-disk stars in the solar cylinder. Our model results are plotted as solid curves for the thin (blue) and thick (red)  disks. 

In our standard model, the MD  of thin disk (panel A1) peaks at [Fe/H]$\sim 0$ and compares fairly  well with the APOGEE sample except a slightly extended tail to the lower metallicities. The introduction of SFBs  pushes the peak metallicity of the MD to lower values  and broadens it significantly (subsequent panels in first row). The largest dispersion of the MD is due to the fact that  a large fraction of the stars is made during, or shortly after, the SFBs, when the Fe abundance increases substantially. 
In the extreme case of 3 local SFB, the first burst 6 Gyr ago triggers a secondary peak at [Fe/H] $\sim$ -0.3 dex (panel D1). It is worth noting that such a double-peaked  distribution, at approximately the same metallicity values, is found in the APOGEE DR17 data in the range 7-9 kpc evaluated at stellar birth radii, as reported in Fig. 4 of  \citet{Spitoni_2024}.  In that same figure, the [Si/Fe] vs [Fe/H] diagram displays two overdensities, at approximately the same positions as in panel D5 of our Fig. \ref{fig:FeH__Hist_Age_XFe}. Taking into account uncertainties in stellar ages, evaluation of stellar birth radii (see Sec. 6 below) and in our radial migration model, we consider the above as an interesting but not conclusive finding.

To summarize this section, we find that the local MDs may be considerably affected by the introduction of recent SFBs and may be used as a constrain regarding the SFB intensity and perhaps other properties (like their extent in space and time). However, the detailed investigation of those effects is beyond the scope of this study.

\subsubsection{Age vs metallicity}
\label{subsubsec:Age_vs_Met}

In the second row of Fig. \ref{fig:FeH__Hist_Age_XFe}, we display age vs metallicity  results for the gas in all radial zones and in the third row  we show the situation for stars present today in the solar cylinder. We compare them with the corresponding data from SDSS and APOGEE obtained in \citet{Feuillet_2019}. We use their results in the zone limited by  7<R<9 kpc and 0<|Z|<0.5 kpc.The behaviour of the average age in the data could be divided into 4 phases strongly correlated with radial migration as discussed in \citet{Kubryk_2015a} and \citet{Prantzos_2023}. Obviously, the first phase of lowest [Fe/H] and highest \afe corresponds to stars older than 9 Gyr, belonging to the thick disk in our models. The relative paucity of stars with ages between 9 and 6 Gyr is as observed in corresponding panel A4 of Fig. 1 and discussed extensively in \cite{Prantzos_2023}. After this transition, the 
average age decreases slowly with metallicity up to about solar [Fe/H], i.e. the present day  metallicity of the local gas and young stars. The subsequent rising of the curve, both in the model (panel A3) and the data of \citet{Feuillet_2019}, is attibuted to the fact that local supersolar metallicity stars originate in the inner, higher metallicity,  disk and need more time on average to radially migrate to the solar neighborhood \citep{Kubryk_2015a,Prantzos_2023}. The last downturn to lower ages observed for the highest metallicities is attributed to a small number of relatively young stars ($\sim$3 Gyr old) formed at Galactocentric distances of R$_G \sim$2 kpc where the metallicity is $\sim$0.4-0.5 dex ; those stars migrated rather rapidly to the solar vicinity, at an average speed of $\sim$2 kpc/Gyr, compared to the typical average speed of $\sim$1 kpc/Gyr.

Considering the relatively large errors in age determination, our standard model presents a fairly good agreement with observational data. On the other hand, all the models with SFBs preserve the 'Down-Up-Down' feature of the relation of average age with metallicity, but the local SFBs make the corresponding slopes steeper (swallower) for local (global) SFBs. Also, local SFBs reduce the minimum average age (around solar [Fe/H] by about a Gyr. In any case, the current uncertainties in age determination make it difficult to conclude anything on the basis of existing observations.

\subsubsection{\afe vs [Fe/H]}
\label{subsubsec:XFe}

In the forth row of Fig. \ref{fig:FeH__Hist_Age_XFe} we show the relation of [Si/Fe] vs [Fe/H] for the gas in all the zones. For the standard case, the  evolution is smooth and there is no overlap among the evolutionary tracks of the various zones. In the two scenarios of local SFBs, the tracks of the intermediate zones (blue) overlap strongly,  due to the significant increase of their [Si/Fe] with increasing [Fe/H]. In those zones,  [Si/Fe] reaches a local maximum near the peak of star formation, and then plunges to a nearly normal level as the Fe of the delayed SNIa of the SFB catches up the excess O released earlier by the CCSN of the SFB. The tracks of zones affected by the local SFBs cross  largely with those of inward zones ( panels B4 and especially D4). Models with global SFBs display similar but less pronounced trends in the 4th row.

The bottom row of Fig. \ref{fig:FeH__Hist_Age_XFe} compares our predictions in the plane  of [Si/Fe] vs [Fe/H] for the solar cylinder. The two sequences of high [$\alpha$/Fe] and low [$\alpha$/Fe] are clearly distinguished in the standard scenario (panel A5), and correspond to the thick and thin disks respectively. As discussed extensively in \citet{Prantzos_2023}, secular evolution can produce such a double-branch behaviour, which is due to the inside-out formation of the disk combined with radial stellar migration and different evolving timescales for the  $\alpha$ elements (originating from short-lived CCSN) and for Fe (from SNIa with long-lived progenitors): very few old stars are formed in the local disk (because of the assumed inside-out formation), thus the locally observed old (and \afe rich) population is brought here by radial migration from the inner disk. This idea is in line with the conclusions of \cite{Buck_2020}
who used hydrodynamical plus N-body simulations to find that the local high \afe sequence  is due to stars formed in the inner disk while the local low \afe sequence corresponds to stars formed over the whole disk ; in both cases, those stars have been distributed across the disk through radial migration. In fact, the model of \citep{Buck_2020} is a kind of "hybrid" model requiring both radial migration to bring locally inner disk stars for the high-alpha sequence and a "wet" merger several Gyr ago in order to dilute the metallicity of the interstellar medium and reproduce the low alpha-sequence. In our model, the high-alpha sequence is obtained in a similar way, but the low-alpha one is naturally obtained by the slow local evolution. The model of
\citet{Buck_2020} apparently implies that without a merger there would only exist the high-alpha sequence in the solar vicinity.

The introduction of SFBs modifies that scheme, by enhancing the young population (as already discussed in the previous section) and, consequently, enhancing the low \afe \ sequence at the expense of the high \afe one. However, the double-branch behaviour of \afe \ vs [Fe/H] is preserved, in general, except the case of panel D5, where the older and strongest  of the two SFBs of \cite{RuizLara_2020} creates an  overdensity bridging the gap between the high and low \afe \ sequences because of the "crossing" of the tracks obtained in panel D4.

In summary, the introduction of SFBs in our baseline models perturbs the double-branch behaviour of the local
[$\alpha$/Fe] vs [Fe/H] relation by a) weakening the high [$\alpha$/Fe] branch  and b) making the two branches join, in the case of a strong and short SFB - as the oldest one of the \citet{RuizLara_2020} study (panel D5  of Fig. \ref{fig:FeH__Hist_Age_XFe}).

\begin{figure*}
	\includegraphics[width=1.\textwidth]{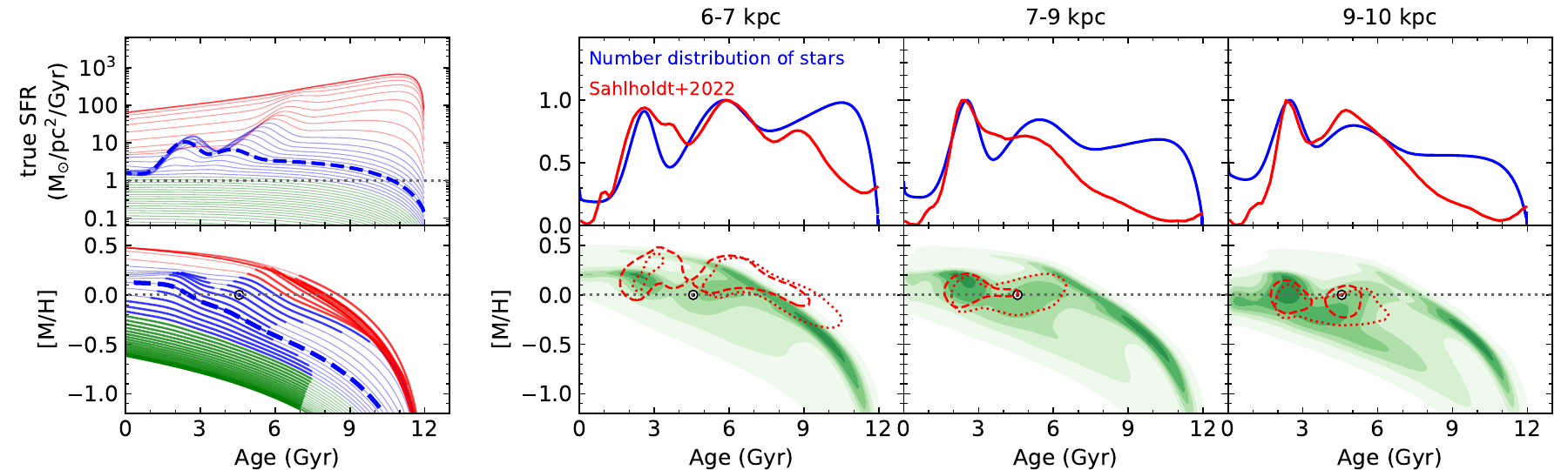}
    \caption{$\it Left$: Same as the second and fourth row of Fig. \ref{fig:Age__SFR_FeH_XFe}, for the evolution of SFR and [M/H] in the propagated SFB model. $\it Right$: Age distributions are displayed in the top panel for different radial ranges of 6-7, 7-9 and 9-10 kpc, corresponding to the equivalent observational subsamples in inner, local and outer bin shown in Fig.6 of \citet{Sahlholdt_2022} ; observations (red curves) are compared to model results (blue curves). In the lower panels,  age-metallicity relations  of the model (green isodensity contours)  are compared to  half-maximum density contours  from observations of \citet{Sahlholdt_2022}  (dotted and dashed red curves indicate  the high-$z_{\textup{max}}$ and low-$z_{\textup{max}}$ part of observations in each radial range). 
    }
    \label{fig:Age__Hist_FeH_XFe_Sahl}
\end{figure*}

\section{Propagated SFB}
\label{sec:Prop_SFB}


In order to reproduce the observations of \citet{Sahlholdt_2022}, which concern not only the local disk but also the inner and outer ones (see the three lower panels of Fig. \ref{fig:SFH_OBS}) we adjust the parameters of Eq. \ref{eq:insideoutSFB}. We adopt two SFBs:  the older SFB is described with center radius $\mu_R$ = 5.5 kpc and propagates more slowly outwards  $t_R$ = 0.42 Gyr/kpc, while the younger one is centered on $\mu_R$ = 7 kpc and propagates faster, with  $t_R$ = 0.16 Gyr/kpc.  The second one is weaker and affects a smaller radial range.  In the top left panel of Fig. \ref{fig:Age__Hist_FeH_XFe_Sahl} we display the true star formation histories of all the radial zones of the model, showing the 2 propagated SFBs.

As shown in the right top panel of Fig. \ref{fig:Age__Hist_FeH_XFe_Sahl}, 
this arrangement allows us to reproduce roughly the  age distributions of inner, local and outer subsamples from \citet{Sahlholdt_2022}. Our propagated SFB scheme allows us to reproduce the different times of the older SFB peaks observed in the various zones, as indicated. Although the positions and intensities of both SFBs are rather well reproduced, the earliest parts of the number distributions of stars in all zones seem to be overproduced, but this is most probably due to the magnitude limited sample adopted in \citet{Sahlholdt_2022}, which disfavours older objects.

The bottom rows of Fig. \ref{fig:Age__Hist_FeH_XFe_Sahl} display the age-metallicity relation of our propagated model.
On the left we show that relation in the gas for all the zones of the model. As already seen in the previous sections, the introduction of SFBs produces in the affected zones a sudden increase of metallicity soon after, with Fe released initially from CCSN and on a longer timescales from SNIa. 
We shall see in the next section how this behaviour may affect the evolution of radial abundance gradients.

In the right bottom panels of Fig. \ref{fig:Age__Hist_FeH_XFe_Sahl}  we display the isocontours of the number densities of stars currently found in the three radial zones (inner, local and outer) in the age-metallicity plane. We superpose the half-maximum contour curves for stars of high-$z_{\textup{max}}$ (dotted) and low-$z_{\textup{max}}$ (dashed) from \citet{Sahlholdt_2022}. Here we use the same recipe as \citet{Sahlholdt_2022} - adapted from \citet{Salaris_1993} - to define the metallicity M/H as:
\begin{equation}
\rm[M/H] = [Fe/H] + \log(0.638 x 10^{[\alpha/Fe]} + 0.362)
\label{eq:MH}
\end{equation}
and we use Si  as proxy for $\alpha$ elements to calculate the model [$\alpha$/Fe]. 

We find that the overdensities obtained in our model correspond fairly well to those of \citet{Sahlholdt_2022}, both in metallicity and age, especially if one takes into account the uncertainties in the evaluation of the latter.    In the inner disk (5-7 kpc) , the intermediate age population (corresponding to the older SFB) is connected to the late thick disk stars and shares almost the same average metallicity with the younger SFB, around [Fe/H]=+0.2. In the local disk (7-9 kpc), the average [M/H] of two corresponding populations decreases from $\sim 0.2$ to near solar - again in excellent agreement with the data -  and there is no connection to the thick disk population.
Finally, in the outer disk, there are still two overdensities, but only the younger one has the correct age and metallicity, while  the older one is less metallic by $\sim$0.2 dex compared to the \citet{Sahlholdt_2022} results. Overall, our propagated SFB model manages to reproduce the position of the two overdensities in the age-metallicity plane, as derived from those observations, in both the inner and local disk, but fails to reproduce the older one in the outer disk.

\section{SFBs and the evolution of the metallicity gradient at birth radii}
\label{sec:met_grad_sfb}

In this section, we discuss the impact of SFBs on the abundance profile of the Galactic disk, and on the possibility of the "reconstruction"  of its past evolution through local observations.

The evolution of the abundance profile plays a key role in our understanding of the MW disc. Most studies predict a flattening of the abundance profile of the disc gas with time as a natural result of the inside-out formation of the disk 
\cite [e.g.][]{Ferrini_1994,Boissier_1999,Hou_2000,Fu_2009, Kobayashi_2011, Pilkington_2012,
Gibson_2013,Kubryk_2015a,Molla_2017,Tissera_2017,Molla_2019}, while others  predict the opposite effect  \citep[e.g.][]{Diaz_1984,Chiappini_1997,Schoenrich_2009,Mott_2013,Grisoni_2018,Vincenzo_2020} due e.g. to the assumption that the whole early disk was rapidly formed and fully mixed or to late dilution of the outer disk with metal poor gas from mergers.

If stars maintained their radial position at birth during galactic evolution, then a good knowledge of their ages and their Galactocentric distances would allow one to reconstruct the evolution of the gaseous abundance profile. For many years, objects covering a substantial age range (several Gyr) and bright enough to be seen in a broad range of Galactocentric distances, like planetary nebulae and open clusters, were used to monitor the evolution of the Galactic abundance gradient \citep[e.g.][and references therein RECIO]{Maciel_2006,Maciel_2007,Magrini_2017,Magrini_2023}

However, as noticed in e.g. \cite{Minchev_2013,Minchev_2018, Kubryk_2015b} the stellar abundance profile of the disc has been strongly altered by radial migration (see e.g. Fig. 6 of the latter work). This idea is now confirmed \cite[see][]{Anders_2023}, and makes it difficult to trace back the evolution of the true abundance gradient in the gas of the disc (that is, at the birth radii of stars observed today).

Based on an idea of \cite{Minchev_2018}, \cite{LuMinchev_2022} attempted recently to solve that problem. They explored the results of different suites of cosmological hydrodynamical simulations of Milky-Way-like galaxies from \cite{LuBuck_2022} who found a tight anticorrelation between the ISM gradient $\nabla$[Fe/H] (in dex/kpc) at a certain look-back time and the locally observed 5\% - 95\%  percentile range of metallicities $\Delta$[Fe/H] (in dex) of the sample stars with the corresponding age. \cite{LuMinchev_2022} applied this finding to a sample of $\sim$80000 subgiant stars of LAMOST DR7 presently located in the Galactocentric distance range R$_G$=6 to 12 kpc, with an average age and metallicity uncertainty as small as 0.32 Gy and 0.03 dex, respectively. They adopted a present day value of -0.07 dex/kpc for the present day gradient and they inferred the value of the central disk metallicity [Fe/H](R$_b$=0) from the upper envelope of the observed age-metallicity relation. Then they fixed the coefficient of the linearity between the range $\Delta$[Fe/H] and the gradient
 at -0.15 dex/kpc, because this value allows them to reproduce better the statistics (guiding radius distribution) of their local stellar sample. Thus,  they found a steepening of the [Fe/H] gradient at birth radius with age, up to 8-9 Gy ago; at that time the trend is inversed and the gradient flattens at the oldest ages. The authors associated that feature to the last major merger, namely the Gaia Sausage/Enceladus event \citep{Belokurov_2018,Helmi_2018}.

More recently, \citet{Ratcliffe_2023} followed a similar method as \citet{LuMinchev_2022} and recovered the metallicity gradient based on $\sim 140000$ APOGEE DR17 red-giant disk stars combined with ages from \citet{Anders_2023}. They found a similar trend as \cite{LuMinchev_2022}, but with two additional fluctuations in the temporal behaviour of the gradient, which they attributed to recent star formation bursts, since the fluctuations occur at approximately the times of passages of Sagittarius dwarf galaxy.

\begin{figure}[t!]
	\includegraphics[width=0.49\textwidth]{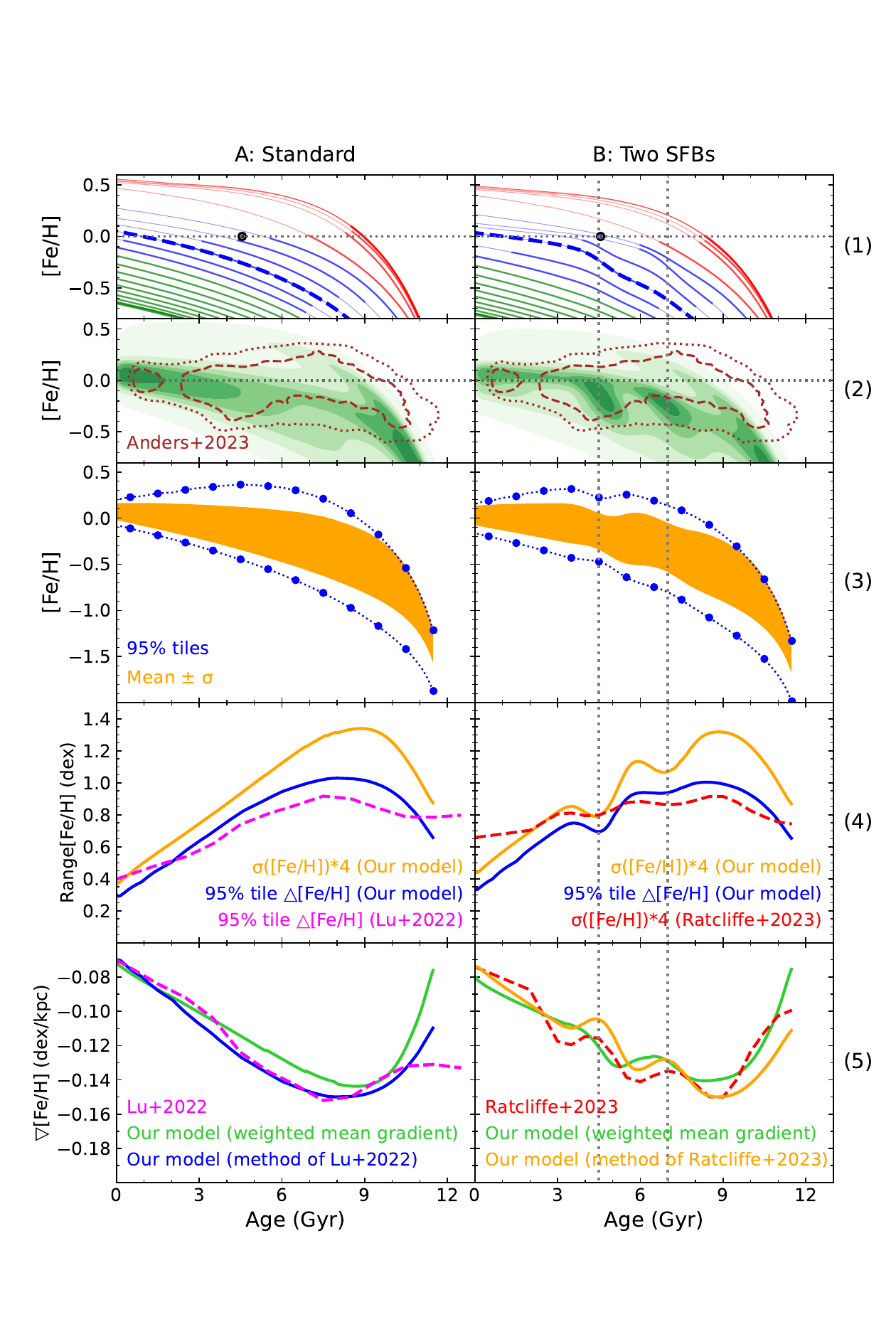}
    \caption{Age-metallicity relations and gradient evolutions in our standard model (left column, compared to \citet{LuMinchev_2022})  and in the 2-SFB model (right column, compared to \citet{Ratcliffe_2023}). $\it Row\ 1$: Gaseous metallicity every 2-zones of the model.
    $\it Row\ 2$: Density isocontours for the solar vicinity in the model (green shaded, as in Fig. \ref{fig:Age__SFR_FeH_XFe}) and in \cite{Anders_2023} (brown); the introduction of a SFB at 4.5 Gyr and 7 Gyr produces two local overdensities, as in the data.
    {\it Row  3}: Age-metallicity relation in solar vicinity with metallicity ranges indicated by mean$\pm$1 $\sigma$ values (orange shaded aerea) and 95\%-tiles (between the blue dotted curves featuring larger points every Gyr).
    $\it Row\ 4$: Metallicity ranges  $\Delta$[Fe/H] (blue and orange curves are derived from the data of same colour in the previous panels) and observationally inferred by \citep[dashed magenta, left panel]{LuMinchev_2022} and  \citep[dashed red, right panel]{Ratcliffe_2023}. 
    $\it Row\ 5$: Evolution of the abundance gradient, in observations from \citep[dashed magenta, left panel]{LuMinchev_2022} and  \citep[dashed red, right panel]{Ratcliffe_2023}, and in our model obtained either with our method of mass-weighted gradient (solid green curves in both panels, see text) or with the method of \citep[solid blue, left panel]{LuMinchev_2022} and \citep[solid orange, right panel]{Ratcliffe_2023}. }
    \label{fig:Age__FeH_Rad}
\end{figure}

In a previous work \citep{Prantzos_2023}, we explored the idea of \cite{LuMinchev_2022} in order to infer the evolution of the stellar abundance gradient at birth radius in the model of the Milky Way from present-day observations of the age and metallicity of local stars.  We showed that a qualitatively similar behaviour as in \cite{LuMinchev_2022} can be obtained, albeit for different reasons. We present the method below, in more detail than in \cite{Prantzos_2023}, and then we discuss our results for the baseline (no SFB) model and  with SFBs.

\subsection{The method}
\label{sub:TheMethod}

\begin{figure}
	\includegraphics[width=0.49\textwidth]{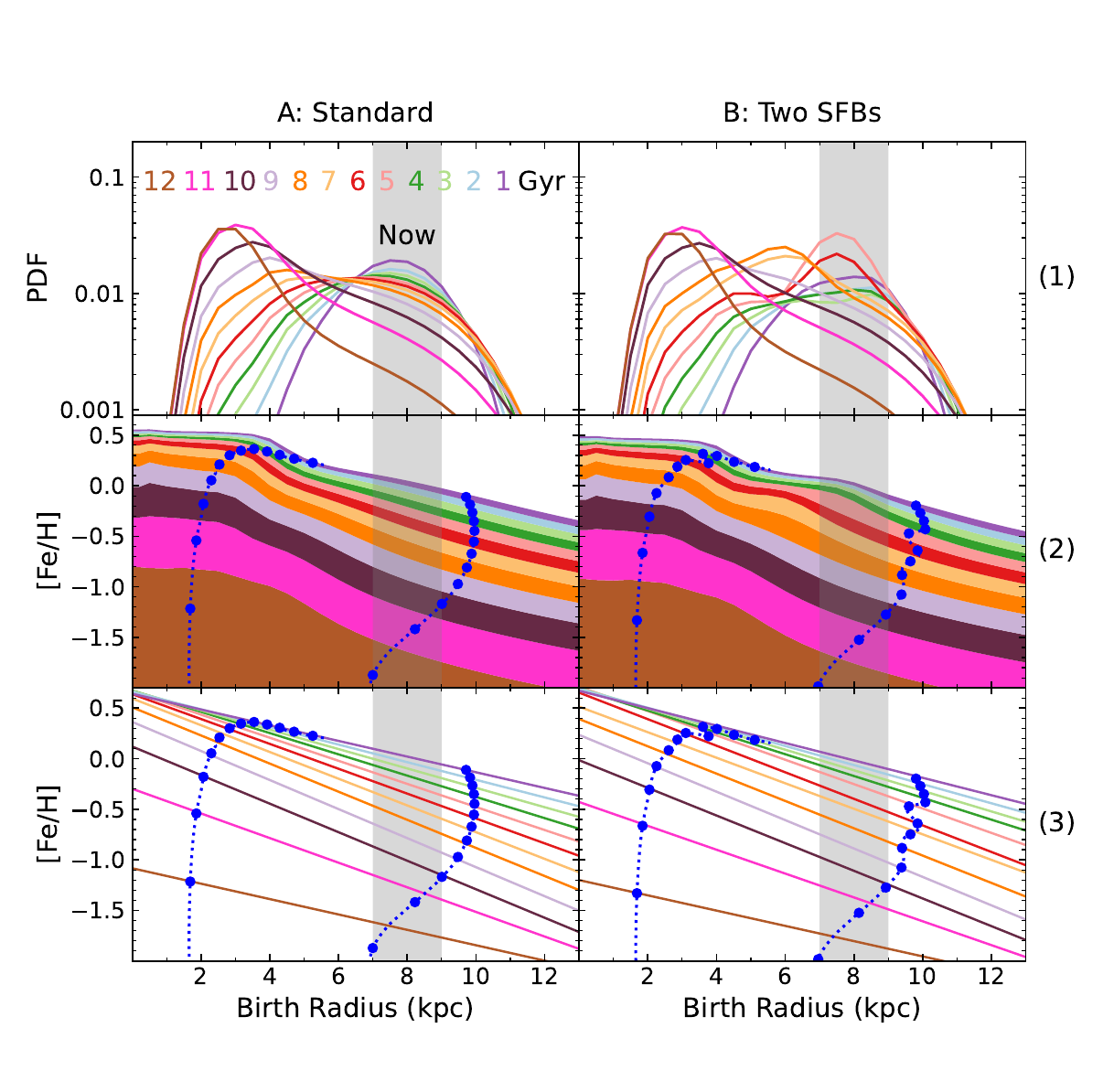}
    \caption{$\it Top$: Distribution of birth radii $\rm R_b$ of stars currently found in the solar vicinity  (grey shaded area in all panels, extending from 7-9 kpc) for stars of different ages (in 1-Gyr wide bins, color coded by age), as predicted by our standard (left) and 2-SFB (right) models. $\it Middle$: Radial metallicity profile of each mono-age population. The blue dots indicate boundaries including 95\%-tiles of model local stars in the $\rm R_b$ distributions of the top panel.  $\it Bottom$: The  dashed lines represent the mass weighted (according to the distributions of the top panel) abundance gradient of each mono-age population.}   
    \label{fig:Rad__PDF_FeH}
\end{figure}

In Figs \ref{fig:Age__FeH_Rad} and \ref{fig:Rad__PDF_FeH} we present our method in detail, step by step.  Fig \ref{fig:Age__FeH_Rad} displays results as function of age and Fig. \ref{fig:Rad__PDF_FeH} as function of birth radius R$_b$. The latter figure explicates some of the steps undertaken in the former and we shall move back and forth between the two during the presentation of the method. In both figures, the left column (A) displays our standard model (without SFBs) and the right one (B) the model with two SFBs. The former model is compared to the results of \cite{LuMinchev_2022}, as already done in \cite{Prantzos_2023} while the latter to the results of \cite{Ratcliffe_2023}. In model (B) with SFBs, the epochs of the two star formation episodes (at 4.5 and 7 Gyr) are indicated with vertical dotted lines through all the panels.

In the top panels of Fig \ref{fig:Age__FeH_Rad}, the evolution of [Fe/H] in the gas of various zones is displayed, with the same color coding as in Fig. \ref{fig:Age__SFR_FeH_XFe}. In model (B) on the right, the SFB results in a sharp increase of metallicity in several zones.
In panels A2 and B2 of Fig. \ref{fig:Age__FeH_Rad} we show the age-metallicity relation of stars presently found in the solar vicinity, compared to the isodensity contours found in the local APOGEE sample by \cite{Anders_2023}. Our model (B) reproduces the overdensity at $\sim$4.5 Gyr found in that work and creates a metal-poorer overdensity at $\sim$7 Gyr, because we kept the radial positions of the 2 SFBs the same as in the propagated SFBs model of the previous section but we adjusted  their temporal positions of our model according to the \cite{Anders_2023} data.

In panels A3 and B3 we display the same local age-metallicity relation, delimited by either the  $\pm$1 $\sigma$ limits (orange shaded aerea) or the 95\%tiles (the two blue dotted curves). The formerlimits have been used in the works of \cite{LuMinchev_2022} and \cite{Prantzos_2023} and we adopt them here for consistency with our previous work, while the latter are adopted in \cite{Ratcliffe_2023} and we compare with that work our model (B) on the right column. In all those cases, it is clear that at late times the range of local stellar metallicities is small (because of little time available for radial migration), it increases with age (because of more time available for radial migration) and then becomes narrow again (because the older local stars are formed in the inner disk, which has evolved in short timescale with high star formation efficiency and in a highly turbulent environment as assumed in our model, resulting in fairly similar early age-metallicity trajectories). The same behaviour is generally obtained on panel B3, but its delimiting curves display a "wiggly" behaviour, with their width temporally compressed during each episode of enhanced star formation.

In panels A4 and B4, the behaviour of the metallicity range ${\rm\Delta[Fe/H]=[Fe/H]_{max}(\tau)-[Fe/H]_{min}({\tau})}$, analysed in the previous paragraph, appears respectively in solid curves for our models (blue for the 95\% tiles limits  and orange for the 1$\sigma$ limits) and in dashed curves for the observational-based inferences of \citet{LuMinchev_2022} (magenta dashed on panel A4) or \citet{Ratcliffe_2023} (red dashed on panel B4). All the curves are first increasing with age, then go through a maximum around 8-9 Gyr and decrease again, for the reasons exposed in the previous paragraph

Up to this point (the 4th rows in Fig. \ref{fig:Age__FeH_Rad}), our method exactly follows that of \citet{LuMinchev_2022} and \citet{Ratcliffe_2023}, providing the metallicity range $\Delta$[Fe/H] of stars observed locally as function of age. However, our approach diverges from here onwards, because we know the gas metallicity profiles (i.e. at the birth radius of the locally observed stars) in our model  throughout the Galactic history, while  \citet{LuMinchev_2022} and \citet{Ratcliffe_2023} have to infer them, making a couple of assumptions, as discussed above. One of their main assumptions (to be discussed below) is that {\it the metallicity profile is characterized at any time by a unique slope over the whole radial range. }

In our case, the evolution of the abundance gradient at birth place inferred through a sample of stars presently found in the solar vicinity is obtained using a slightly different method from \citet{Prantzos_2023}. The method is illustrated in Fig. \ref{fig:Rad__PDF_FeH}, again for the standard model (left column) and the 2 SFB model (right column). In the top panels of Fig. \ref{fig:Rad__PDF_FeH}, we present the distributions of birth radii $R_b$ for stars currently present locally (i.e. within the grey shaded aerea), color coded as function of stellar ages (with 1 Gyr age bins). It can be seen that the older stars currently present in the solar vicinity,  have been formed mostly in the inner disk. The distributions of birth radii move towards solar vicinity with time, because younger stars have less time to migrate.  In panel B1 of that figure, itcan be seen that the introduction of SFBs modifies the birth place distributions at intermediate ages.
In panels A2 and B2, we display the gaseous metallicity profiles of the Galactic disk for given age bin (1 Gyr wide), color coded as  indicated in panel A1. The 2 blue dotted curves delimit  the R$_b$ range corresponding to the 95\% tiles of the metallicity ranges of local stars at each age bin (as shown in panels A3 and B3 of Fig. \ref{fig:Age__FeH_Rad}. One immediately sees that the gas profile inside the region $\Delta$ R$_b$ between those two delimiting radii has indeed a single-slope at late times but not so at early times, since the early metallicity profiles in the inner disk are flat (following our assumption of an early period of very rapid star formation and highly turbulent gas). This indicates that the assumption made by \cite{LuMinchev_2022} and \cite{Ratcliffe_2023} (a single slope of the metallicity profile over the whole radial range at any given time) is not  applicable, at least in the case of our model. Moreover, the introduction of recent SFBs also modifies  the recent metallicity profiles which deviate from a single exponential even at intermediate times and Galactocentric radii.

We proceed then, as shown in the bottom panels of Fig. \ref{fig:Rad__PDF_FeH}, to calculate a mass-weighted gradient for each age bin $\tau$, {\it assuming that there is a linear relationship between }[Fe/H] and $\rm R_b$. 

The result of such a fitting of the abundance profile at birth radius assuming a single slope
appears in the bottom panels of Fig. \ref{fig:Rad__PDF_FeH}. The oldest profile (lowest curve at age of 11 Gyr ) is flatter than the subsequent ones, being dominated by the stars of the inner disk; it becomes more and more steeper until the age of 8-9 Gyr ago, when it starts getting flatter again. The introduction of two SFBs (column B) modifies the distributions of the birth radii (panel B1) and the abundance profiles at the epoch of the SFBs (panel B2). We note  that the method of mass-weighted gradient adopted here gives slightly different results that the simpler method adopted in \cite{Prantzos_2023}, where the gradient was calculated simply as $\Delta$[Fe/H]/$\Delta$R$_b$.

\subsection{Results}
\label{sub:Results_Evol_Grad}

The results of Fig. \ref{fig:Rad__PDF_FeH} help to understand the  evolution of the abundance gradient at birth radius inferred from  local observations in our model. This evolution appears in the bottom panels of Fig. \ref{fig:Age__FeH_Rad} where they are compared with the results of \cite{LuMinchev_2022} (A5) and \cite{Ratcliffe_2023} (B5). We present results derived from two different methods: the one discussed in the previous paragraph (green in both panels), and by  adopting the same method as \cite{LuMinchev_2022} and \cite{Ratcliffe_2023}, respectively (blue in A5 and orange in B5). 

The evolution of the gradient obtained in \cite{LuMinchev_2022} is well reproduced in our standard model (A)  both with  our mass-weighted gradient and  with the method of \citet{LuMinchev_2022}, as can be seen by comparing the solid  green and blue curves with the purple dashed curve of the data in panel A5 of  Fig. \ref{fig:Age__FeH_Rad}. The 3 left panels of Fig. \ref{fig:Rad__PDF_FeH} help to understand that evolution. During the earliest period, the local stars we observed today come essentially from the innermost regions (R$_G$<3 kpc, see panel A1 left in Fig. \ref{fig:Rad__PDF_FeH}) where the metallicity profile was flat (A2  panel in same figure), providing a small slope (in absolute value)  for the corresponding abundance gradient (panel A3 in the same figure). Progressively, the birth place distribution of presently local stars is shifted to intermediate regions (3-5 kpc, panel A1 left) which were characterized by a steep abundance profile, thus increasing the absolute value of the negative abundance gradient. This increase stops around the age of 9 Gyr, when the birth place distributions start being dominated by the regions around 5-8 kpc, where the abundance profile becomes progressively flatter and thus the abundance gradient starts decreasing in absolute value. Thus, the maximum in absolute value observed around 9 Gyr corresponds to the end of a period of intense activity which had originally produced a flat profile in the inner disk leaving the outer disk with a very steep profile progressively dominating the local stellar sample. This was also discussed in \citet{Prantzos_2023}, who found that the non-monotonic evolution of the abundance gradient at birth radius results naturally through secular evolution, as several other properties of the local disk.  In contrast, \cite{LuMinchev_2022}  attribute it to the impact of a major merger (GES) on the metallicity profile of the Galaxy 8-10 Gyr ago.

In the case of  model (B), a similar non-monotonic trend as in the baseline  model is obtained, but this time two wiggles are superimposed, resulting from the action of the 2 SFBs.  The effect of the SFBs is to increase rapidly the local metallicity, more in the outer regions than in the inner ones, because metallicity is smaller in the former and thus easier to increase. As a result, the metallicity difference between the inner and outer zones  decreases during the SFBs. In fact, it is the Fe ejected by CCSN which is responsible for that, the Fe of SNIa being released on much longer timescales.  When we adopt the method of  \cite{Ratcliffe_2023}, namely by using the "mirror image" of the metallicity range and adjusting the present day and minimum values, our result is in fair agreement with the results of that work (solid orange vs dashed red curves, respectively, in panel B5 of  Fig. \ref{fig:Age__FeH_Rad}). 

On the other hand, our own method (using the mass-weighted gradient at birth place for all stars within a 1 Gyr-wide age bin) identifies the impact of the older SFB at age of 7 Gyr with  a slight time-delay by about 0.3 Gyr,  probably because it integrates partially the impact of SNIa originating by that SFB. However, it fails to identify clearly the impact of the younger SFB of age 4.5 Gyr: only a change in the slope of the evolution of the gradient, again with a slight time-delay, indicates the existence of some star formation episode at that time. The difference with respect to the case of the earlier SFB is due to the fact that at early times the mass weighted gradient is determined by the stars of restricted regions in the inner disk, while at late time it is smoothed over a large region of birth radii.  

\subsection{Discussion}

Comparing the results obtained with the two different methods of analysis, we conclude that the idea of \cite{LuMinchev_2022} may be applicable if the metallicity profile has at any moment a unique slope over the whole range of birth radii, but it leads to umbiguous results if the profile is a multi-slope one, as naturally obtained in the case of localized star formation episodes.

We note that in a recent work \cite{Ratcliffe_2024} analysed Milky Way/Andromeda analogues from the TNG50 simulation to study the possibility to recover the behaviour of the metallicity profile at R$_{birth}$ as function of radius in a variety of galaxies. They find that  the true central metallicity is unrepresentative of the genuine disk [Fe/H] profile and suggest  to use a projected central metallicity instead. They conclude that "Looking in detail at a Milky
Way-like galaxy... we expect to be able to recover R$_{birth}$ to within 1 kpc for the Milky Way disk.". Similarly, \cite{Lu_2024} investigate the possibility of inferring birth radii for external galaxies such as the LMC using the NIHAO cosmological zoom-in simulations. They find that it is theoretically possible to do so, with a $\sim$25 per cent median uncertainty for individual stars, provided several conditions are fulfilled. However, the possibility of a gradient with a non-unique slope over the whole redial range is not considered in those  works.

\section{SFB driven by gas dilution of the outer disk}
\label{sec:SFB_infall}


In the previous sections we considered SFBs induced by enhancement of the star formation efficiency, due for instance to perturbations from the passage of a nearby satellite galaxy. However, episodes of star formation may also be induced by events of strong gas accretion. Such events have a double effect: they first dilute the local gas metallicity and subsequently they increase it strongly, through the resulting strong star formation due to the larger gas reservoir; the latter effect will be  particularly important if at the same time an enhanced star formation efficiency is assumed.

Such effects have been explored in some detail in \cite{Johnson_2020} with 1-zone GCE models, without any connection to specific observational data. In a subsequent paper \citet{Johnson_2021} used a multi-ring model with parametrized stellar radial migration  and applied it to the study of the Milky Way. Within the framework of that model, they explored the consequences of a late and broad SFB, like the one of \cite{Mor_2019} on the present-day abundance gradients of Fe and O, but they provided no information on the impact of those SFBs on evolution of the gradients.

The cosmological simulations of MW-mass galaxies by \citet{Buck_2023}  suggest that metal-poor cold gas from an early massive merger could  dilute the metallicity in the outer disk regions and steepen an otherwise flat metallicity gas profile; this supports the \citet{LuMinchev_2022} interpretation of their observed gradient inversion as due to a major merger 8-9 Gyr ago, perhaps related to the thick-thin disk transition epoch. 
In analogous way, \citet{Ratcliffe_2023} attributed the two more recent fluctuations in the evolution of the abundance gradient to dilution episodes from gas accretion in the disk outskirts, caused by passages of the Sgr galaxy which left behind substantial amounts of low metallicity gas.

 In this section, we explore the consequences of those ideas by simulating two recent episodes  of pristine (zero metallicity) infalling gas. 
We assume  episodes  of gaussian form, with maxima at ages of  6.3 Gyr and 3.7 Gyr ago, respectively,  and standard deviations of 0.5 Gyr each. Those infall episodes are assumed to occur in the outer disk,  and their intensity decreases in the inner radii (it becomes null at 6 kpc). The time-integrated mass of accreted gas
in each ring is normalized as to be the same with the standard case, in order to ensure the same total mass for the Galactic disk.

\begin{figure}[t]
    \centering
	\includegraphics[width=0.49\textwidth]{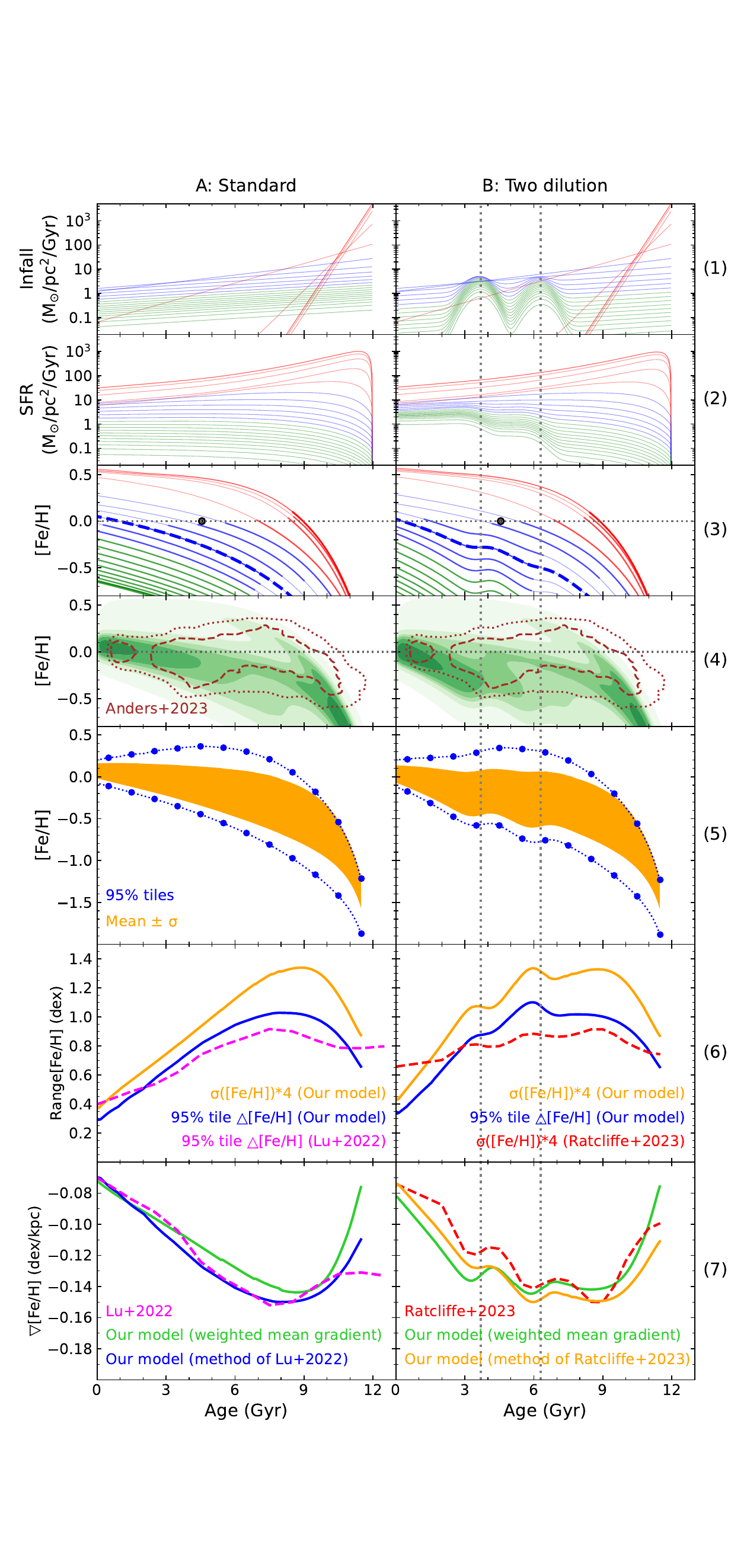}
    \caption{Comparative evolution of several quantities for the baseline model (column A on the left) and the case of two recent infall episodes in the outer disk (column B on the right). {\it Top}: The infall rate in various radial zones, colour coded as in previous figures (red: inner zones, blue: intermediate, green: outer; thick dashed: 8 kpc). {\it Second from top}: Corresponding local star formation rates. The subsequent five rows of panels are the same as in Fig. \ref{fig:Age__FeH_Rad}}.
    \label{fig:Age__Infall_SFR_FeH_Rad}
\end{figure}

The comparison of the standard model and the one with two episodes of gas accretion are displayed in columns A and B, respectively, of Fig. \ref{fig:Age__Infall_SFR_FeH_Rad}. In the first row (panel B1) appear the two infall episodes, which are accompanied with a small time delay by enhanced star formation (panel B2). The late increase of gass mass in the outer disk reduces the difference of SFR between the inner and outer zones, when compared to the standard case (panel A2).  In the subsequent rows of Fig.\ref{fig:Age__Infall_SFR_FeH_Rad}, we show the same results as Fig. \ref{fig:Age__FeH_Rad}. In panel B3, [Fe/H] in the gas of the outer regions is diluted by the additional primordial gas while the inner regions, unaffected by dilution,  keep their steadily increasing trend of metallicity. Consequently, the metallicity range  of stars found locally today increases during the dilution process (panels B5 and B6) and contracts right after that episode, because of the rapid  Fe production through the enhanced star formation. This translates into a steepening of the Fe gradient followed by a flattening, and  the same thing happens after the second dilution episode  (panel B7).

The comparison to the data of \cite{Ratcliffe_2023} in panel B7 is quite satisfactory, whether the gradient is evaluated through the mirror image of the range of [Fe/H] (orange curve) or with the weighted mean method (green curve) although the two methods do not provide exactly the same results. We emphasize that the oldest wiggle in panel B7 (at an age of 8-9 Gyr) is not due to any episode of SFR or gas accretion but simply to the migration of old stars from the inner regions, as it appears in panel A7 and is discussed in the previous section.

A careful comparison between Figs. \ref{fig:Age__FeH_Rad} and \ref{fig:Age__Infall_SFR_FeH_Rad} reveals an interesting feature. In the former case, the time of the SF episode coincides with the maxima of the observationally inferred wiggles (see the vertical dashed curves throughout the figures). This is due to the fact that the SFBs in the outer disk enhance the metallicity in that region, pushing it closer to the metallicity of the inner one; as a result, the range of metallicities $\Delta$\afe \ of mono-age populations found in the solar neighborhood today is reduced and in its mirror image, corresponding to the metallicity gradient, is increased, i.e. a maximum is obtained the time of the SF episode (panels B4 and B5 in Fig. \ref{fig:Age__FeH_Rad}. On the other hand, the infall episodes in the outer disk dilute the local metallicity and increase its difference with the metallicty in the inner regions; as a result, the range of metallicities $\Delta$\afe \ of mono-age populations found in the solar neighborhood today increases and in its mirror image  a minimum is obtained (panels B5, B6 and B7 in Fig. \ref{fig:Age__Infall_SFR_FeH_Rad}. We think that this property may help in the future, through improved age derminations, to better probe the nature of the interaction that produced the wiggles: enhanced SF efficiency in the first case vs infall episode in the second.

\section{Discussion}
\label{sec:Discuss}

As mentioned in Sec. \ref{subsec:Ruiz}, \citet{RuizLara_2020} found star formation episodes $\sim$ 5.7, 1.9 and 1 Gyr ago, and related them to pericentric passages of Sgr, which is commonly thought to be a highly possible perturber of MW in the past \citep{Penarrubia_2010,Gomez_2013}. This view is in agreement with the simulations of the CLUE (Constrained Local UniversE) project of\citet{DiCintio_2021}, where the host galaxies of infalling  satellites show an enhancement of their star formation at pericenter passage. However,  the authors emphasize that the this SF enhancement depends on several factors, like the amount of gas of the perturber at pericenter and its distance to the host: indeed, at very close passages, the gas of the perturber may be removed by ram pressure. In the  High-resolution Environmental Simulations of  the Local Group,  which concern early mergers \cite{Khoperskov_2023} find  a correlation between mergers and close pericentric passages of massive satellites and bursts of the star formation in the in situ component of Milky Way type galaxies. 

Also, 
\citet{Annem_2024}
 made hydrodynamical simulations of interactions of the MW-type galaxies with massive satellites, both gas-poor  and gas-rich ones.
on quasi-polar orbits. They find a substantial increase in the SFR during
 close passages of gas-poor satellites. Such passages cause the formation of low-metallicity stars in the host, since the gas infall from the satellite results in the dilution of the local mean stellar metallicity of the host.
They suggest that star formation burst at $\sim6$ Gyr like \citet{RuizLara_2020} could be attained through close passages (less than 20 kpc) of Sgr progenitor with subsequent non-negligible mass loss.

\section{Summary}
\label{sec:summary}
In this work we explored with a semi-analytical model including radial migration the consequences of recent star-formation episodes - suggested by both observations and hydrodynamical simulations - on some chemical observables of the Milky Way disk. We considered both local and global SF episodes and in all cases we used as key constrain the local SF history (in the solar neighborood) inferred from recent observations.  We focused, in particular, on the resulting evolution of metallicity and \afe \ ratios and of the metallicity gradient of stars at birth place.

We find that SF episodes induced by a temporary increase of the SF efficiency result in overdensities in the phase-space of various quantities, e.g. age vs metallicity, or \afe \ vs metallicity. The also produce a rapid increase of both metallicity and \afe \ ratios. The increase is more pronounced in the case of local than global SF episodes, under the constrain that both cases reproduce the observationally inferred local SF histories. They are also more pronounced for shorter SF episodes, like those reported by \cite{RuizLara_2020}, than long lasting ones, as the one inferred in \cite{Mor_2019}. In all cases, the local metallicity distribution is broadened with respect to the "standard" model with no SFBs, while the average age metallicity 
relation of the local stellar population is distorted to a significant degree. Moreover, the birth place of the Sun has to move a little inwards (about 0.5 to 1.5 kpc) than in the case of our "standard" model  (without SF episodes) presented in \cite{Prantzos_2023}.

We find that such SF episodes may increase the number of "young \afe-rich stars" in the Galaxy, thus alleviating - but definitely not solving - that problem, first noticed by \cite{Chiappini_2015} and \cite{Martig_2015}. Mergers of old \afe-rich stars of the thick disk, becoming more massive and therefore appearing younger, is the current popular explanation of that problem, e.g. \cite{ZhangM_2021,Miglio_2021}. We note, however,  that a fraction of extreme \afe-rich dwarf stars in the GALAH survey has been shown to possess kinematics of thin disk stars and high Li content \citep{Borisov_2022}, putting in question the merger interpretation.

 In  \citep{Prantzos_2023} we found   a non-monotonic behaviour of the local metallicity gradient of stars at their birth radius vs age in our "standard " model: the locally observed abundance profile at birth radius appears steeper at higher ages, until a look-back time of 8-9 Gyr where the trend is inversed. Such a relation was found empirically in \cite{LuMinchev_2022}, who attributed it to an old (8-9 Gyr ago) perturbation of the disk by a major merger like the GES. We showed that in our model this is naturally obtained from the combined effects of the inside-out formation of the disk and radial migration. We also noticed that the method of \cite{LuMinchev_2022} applies when the gaseous abundance profile is described by a unique slope over the whole radial range during the disk evolution, and not by a multi-slope one as in our model.  Here we explore the impact of recent SF episodes on the evolution of the gradient at birth place and we find that such SF episodes can explain the "wiggly" behaviour of the gradient vs age found by \cite{Ratcliffe_2023}, provided that the times of the
SF episodes are adjusted to the times of the wiggle maxima in the latter work (see Sec. \ref{sub:TheMethod} and  Fig. \ref{fig:Age__FeH_Rad}). We also show how a similar trend can obtained through intense infall episodes. An important difference between the two cases is that in the former it is the maxima of the wiggle that correspond to the time of the SF episodes, while in the latter it is the minima that correspond to the time of the infall episodes.

Finally we note that observations of gaseous abundance profiles in resolved galactic disks at high redshifts suggest very little variation of the gradient with look-back time for most of them  \citep{Curti_2020}. This is  in contrast with the significant evolution found in \cite{LuMinchev_2022} and \cite{Ratcliffe_2023} for the MW disk or in our models and in most inside-out schemes for disk formation.  Various effects (turbulence, mergers, interactions) may mix the disk gas, leading to rather flat abundance profiles throughout the disk evolution, while in the absence of such effects disks may indeed develop a steep initial abundance profile flattening with time, see e.g. \cite[][and references therein]{Hemler_2021}. It may well be that our Milky Way belongs to the latter case.

\begin{acknowledgements} 
{This study was conducted  by Tianxiang Chen as a part of his PhD Thesis. Unfortunately, he passed away during the writing of the paper. It is dedicated to his memory.}
\end{acknowledgements} 

\bibliographystyle{aa}
\bibliography{Reference_SFB} 

\end{document}